%% file: main.tex
\documentclass[nofootinbib,pra,twocolumn,showpacs,superscriptaddress,notitlepage,amsmath]{revtex4-2} %,linenumbers

\usepackage{pifont}
\usepackage[colorlinks=true,citecolor=blue,linkcolor=blue,urlcolor=blue]{hyperref}
\usepackage{graphicx}
\usepackage{nicefrac}
\usepackage{blindtext}
\usepackage{float}
\usepackage{multirow} 
\usepackage{xcolor}
\usepackage{colortbl}
\def\ZZ{\mathbb Z}
\def\bs#1{\boldsymbol{#1}}
\newcommand{\cbox}[2]{\vcenter{\hbox{\includegraphics[width=#1em]{#2}}}}
\usepackage{amsmath,amssymb,bm,color}
\usepackage{dsfont}
\usepackage{framed}
\usepackage{braket}
\definecolor{shadecolor}{gray}{0.95}
\usepackage{amsthm}
\usepackage[normalem]{ulem}
\usepackage{siunitx}
\usepackage{tikz}

\usepackage{lineno}
\input{tikz/colorscheme}
\definecolor{red}{RGB}{180, 40, 45}
\definecolor{green}{RGB}{28, 160, 94}
\definecolor{blue}{RGB}{21, 63, 189}

\usepackage{accents}

\newcommand{\vardbtilde}[1]{\tilde{\raisebox{0pt}[0.85\height]{$\tilde{#1}$}}}

\usepackage{wasysym}

\begin{document}

\begin{abstract}
Non-Abelian topological order (TO) is a coveted state of matter with remarkable properties, including quasiparticles that can remember the sequence in which they are exchanged. These anyonic excitations are promising building blocks of fault-tolerant quantum computers. However, despite extensive efforts, non-Abelian TO and its excitations have remained elusive, unlike the simpler quasiparticles or defects in Abelian TO. In this work, we present the first unambiguous realization of non-Abelian TO and demonstrate control of its anyons. Using an adaptive circuit on Quantinuum's H2 trapped-ion quantum processor, we create the ground state wavefunction of $D_4$ TO on a kagome lattice of 27 qubits, with fidelity per site exceeding $98.4\%$. By creating and moving anyons along Borromean rings in spacetime, anyon interferometry detects an intrinsically non-Abelian braiding process. Furthermore, tunneling non-Abelions around a torus creates all 22 ground states, as well as an excited state with a single anyon---a peculiar feature of non-Abelian TO. This work illustrates the counterintuitive nature of non-Abelions and enables their study in quantum devices.
\end{abstract}

\title{Non-Abelian Topological Order and Anyons on a Trapped-Ion Processor}
\author{Mohsin Iqbal}
\thanks{These authors contributed equally to this work.}
\affiliation{Quantinuum, Leopoldstrasse 180, 80804 Munich, Germany}
\author{Nathanan Tantivasadakarn}
\thanks{These authors contributed equally to this work.}
\affiliation{Walter Burke Institute for Theoretical Physics and Department of Physics,
California Institute of Technology, Pasadena, CA 91125, USA}
\author{Ruben Verresen}
\thanks{These authors contributed equally to this work.}
\affiliation{Department of Physics, Harvard University, Cambridge, MA 02138, USA}
\author{Sara L. Campbell}
\affiliation{Quantinuum, 303 S Technology Ct, Broomfield, CO 80021, USA}
\author{Joan M. Dreiling}
\affiliation{Quantinuum, 303 S Technology Ct, Broomfield, CO 80021, USA}
\author{Caroline Figgatt}
\affiliation{Quantinuum, 303 S Technology Ct, Broomfield, CO 80021, USA}
\author{John P. Gaebler}
\affiliation{Quantinuum, 303 S Technology Ct, Broomfield, CO 80021, USA}
\author{Jacob Johansen}
\affiliation{Quantinuum, 303 S Technology Ct, Broomfield, CO 80021, USA}
\author{Michael Mills}
\affiliation{Quantinuum, 303 S Technology Ct, Broomfield, CO 80021, USA}
\author{Steven A. Moses}
\affiliation{Quantinuum, 303 S Technology Ct, Broomfield, CO 80021, USA}
\author{Juan M. Pino}
\affiliation{Quantinuum, 303 S Technology Ct, Broomfield, CO 80021, USA}
\author{Anthony Ransford}
\affiliation{Quantinuum, 303 S Technology Ct, Broomfield, CO 80021, USA}
\author{Mary Rowe}
\affiliation{Quantinuum, 303 S Technology Ct, Broomfield, CO 80021, USA}
\author{Peter Siegfried}
\affiliation{Quantinuum, 303 S Technology Ct, Broomfield, CO 80021, USA}
\author{Russell P. Stutz}
\affiliation{Quantinuum, 303 S Technology Ct, Broomfield, CO 80021, USA}
\author{Michael Foss-Feig}
\affiliation{Quantinuum, 303 S Technology Ct, Broomfield, CO 80021, USA}
\author{Ashvin Vishwanath}
\affiliation{Department of Physics, Harvard University, Cambridge, MA 02138, USA}
\author{Henrik Dreyer}
\affiliation{Quantinuum, Leopoldstrasse 180, 80804 Munich, Germany}

\date{\today}
\maketitle

\section{Introduction}
\begin{figure}[ht]
\centering
\begin{tikzpicture}
\node at (0,0) {\includegraphics[scale=0.5]%[scale=0.3]
%{borrov_arg.pdf}};
{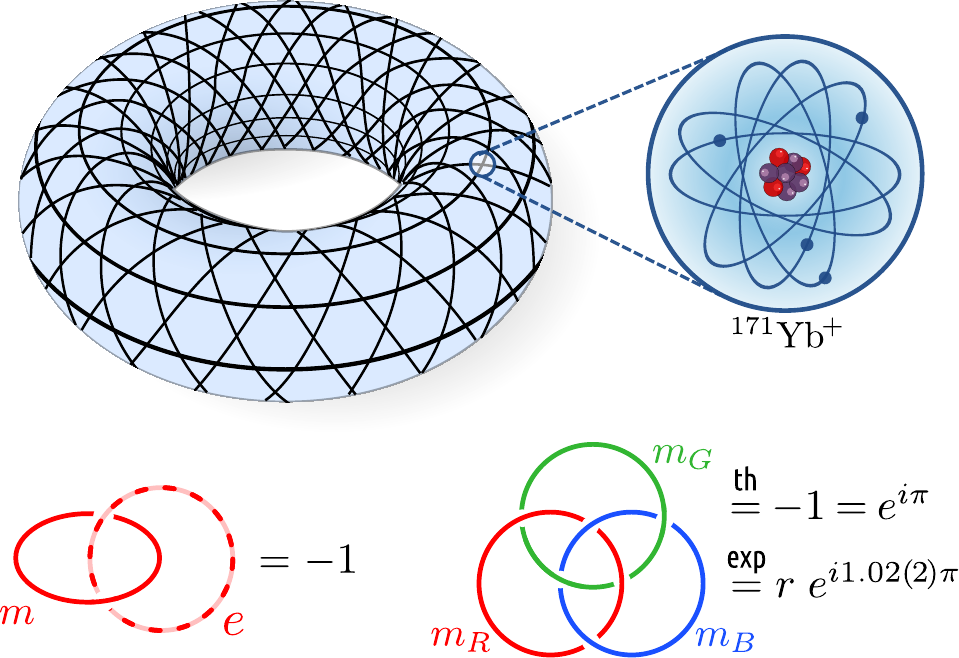}};
\node at (-3.9,2.5) {(a)};
\node at (-3.9,-1.1) {(b)};
\node at (0,-1.1) {(c)};
\end{tikzpicture}
\caption{\textbf{Creating and controlling non-Abelian wavefunctions.} (a) We entangle 27 ions to create the ground and excited states of a Hamiltonian with $D_4$ topological order on a kagome lattice with periodic boundary conditions. (b) Its excitations go beyond Abelian anyons, whose spacetime braiding depends only on pairwise linking, as exemplified by the $e$- and $m$-anyons of the toric code. (c) We create and control non-Abelian anyons $m_{R,G,B}$ which can detect Borromean ring braiding via anyon interferometry; see Fig.~\ref{fig_fuse2red_borromean}b of this work. }
\label{fig_conceptual}
\end{figure}
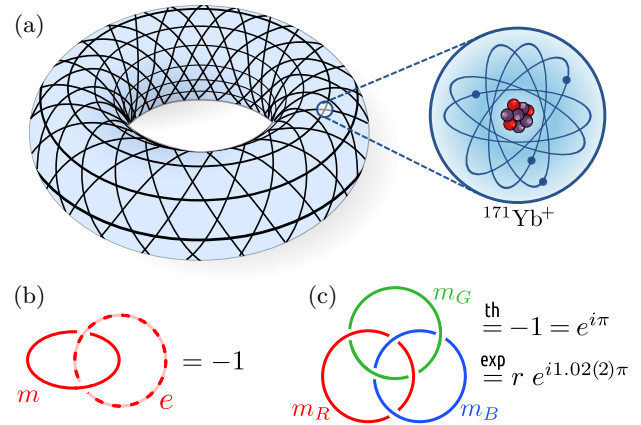

Wavefunctions can exhibit a type of entanglement called `topological order', appearing at the frontiers of condensed matter and high-energy physics and forming the backbone of many proposals for fault-tolerant quantum information processing \cite{wen_quantum_2010}. Such states come in two levels of complexity. The simplest topological wavefunctions are \emph{Abelian}, whose pointlike excitations, called anyons, acquire a phase factor upon braiding one around one another \cite{leinaas_theory_1977,Goldin81,wilczek_quantum_1982} (see Fig.~\ref{fig_conceptual}b). They have been proposed as robust quantum memories \cite{fowler_surface_2012}, and the fractional statistics of Abelian anyons have been verified in certain fractional quantum Hall states \cite{nakamura_direct_2020,bartolomei_fractional_2020}. More recently, the correlations associated to Abelian phases have been probed in a variety of engineered quantum devices~\cite{satzinger_realizing_2021, semeghini_probing_2021,ryan-anderson_implementing_2022, iqbal_topological_2023,Fossfeig23}.

\begin{figure*}[!ht]
	\centering
\includegraphics[width=1.0\textwidth]{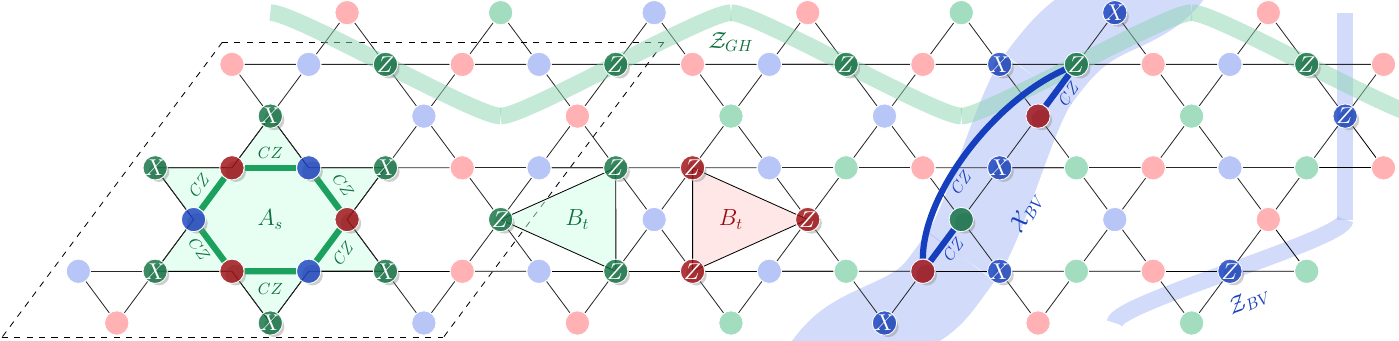}
\caption{\textbf{The non-Abelian $D_4$ model and its logical operators.} We consider the model~(\ref{eq_hamiltonian}) on qubits that live on the vertices of a kagome lattice with periodic boundary conditions. Each kagome star is associated with three local operators: a 12-body star operator $A_s = \prod_{i=1}^6 CZ_{i,i+1} X^{\otimes 6}$ and two 3-body triangle operators $B_t = Z^{\otimes 3}$. It is convenient to choose a vertex coloring and assign a corresponding color to each of the qubits. For each of the three colors and directions along the torus, there are two logical string operators. The logical $\mathcal{Z}$-operators are products of local Pauli-$Z$ acting on all qubits of the respective colour in the chosen direction (\green{$\mathcal{Z}_{GH}$} and \blue{$\mathcal{Z}_{BV}$} highlighted). For the logical $\mathcal{X}$-operator (\blue{$\mathcal{X}_{BV}$} shown) a product of $X$ is applied connecting stars of the the other two colors and decorated with a linear-depth circuit of $CZ$.
More precisely, after choosing a starting point for the vertical string (here, bottom of figure) and direction (here, \blue{blue}$\to$\red{red}$\to$\green{green}), we act with $CZ$ gates connecting every green vertex with preceding red vertices on the path. In the experiment, we implement the system in the black dashed lines containing $3\times3$ stars (27 qubits) and periodic boundary conditions.}
 \label{fig_overview}
\end{figure*}

The situation for \textit{non-Abelian} topological phases is rather different \cite{Goldin85,Moore89,moore_nonAbelions_1991,Wen91}. These more exotic states host excitations called non-Abelian anyons, which come with internal states. Braiding of non-Abelian particles generically effectuates a matrix action on this degenerate manifold. Such braiding is the operating principle of a topological quantum computer~\cite{kitaev_fault-tolerant_2003,nayak_non-Abelian_2008}.
This is associated to robustness to errors and thus defines a coveted goal, as is more generally the controlled realization of a non-Abelian topological phase---bringing their remarkable properties under the experimental spotlight.

Thus far, however, the controlled preparation and braiding of anyons of a non-Abelian topological phase has evaded experiment. To date, the strongest candidate is the fractional quantum Hall state at $\nu = 5/2$ filling~\cite{pan_exact_1999}, which in numerical studies is identified with non-Abelian states.
%the Moore-Read Pfaffian (or anti-Pfaffian) state. 
While thermal Hall measurements corroborate its non-Abelian nature by measuring a fractional chiral central charge $c$~\cite{banerjee_observation_2018}, the integer part of $c$ from the same experiment differs from numerics and remains a puzzle~\cite{ma_fractional_2022}. Furthermore, recent measurement of the non-Abelian statistics of its excitations show promising signs \cite{willett_interference_2023}, although significant challenges remain \cite{feldman_fractional_2021}. While anyons are pointlike, in other settings, the end points of line defects can display similar non-Abelian properties. Well-known examples are Majorana zero modes at the endpoints of $d=1$ topological superconductors~\cite{kitaev_unpaired_2001, aghaee_inas-hybrid_2022} and lattice defects in $d=2$ Abelian topological orders, such as in the toric code \cite{bombin_topological_2010, andersen_observation_2022, xu_digital_2022}. However, to the best of our knowledge, genuine non-Abelian topological order is a necessary condition for existing proposals for universal and topologically-protected quantum computation in two dimensions~\cite{kitaev_fault-tolerant_2003, cui_universal_2015, barkeshli_physical_2015, barkeshli_theory_2013, barkeshli_genons_2013, cong_universal_2017}.

In this work we realise non-Abelian topological order on Quantinuum's H2 trapped-ion quantum processor. This device is based on the quantum charge-coupled device (QCCD) architecture \cite{wineland_1998,kielpinski_2002}, utilizing qubits encoded in the ground-state hyperfine manifold of $^{171}$Yb$^{+}$ ions as described in detail in Ref.~\onlinecite{moses2023race}. The effectively arbitrary connectivity of the platform is used to create an entangled state of 27 ions with $D_4 \cong \mathbb Z_4 \rtimes \mathbb Z_2$ gauge symmetry (see Fig.~\ref{fig_conceptual}a), with the ions assigned to the vertices of a Kagome lattice on a 2D torus. We showcase increasingly exotic properties specific to non-Abelian topological order, including (i) a non-square ground state degeneracy of 22,\footnote{This is in contrast with Abelian topological orders that admit a gapped boundary, which must necessarily have a square ground state degeneracy, cf. section~\ref{methods_nonsquare_GSD}.} (ii) braiding of non-Abelian anyons causing a non-trivial action on the ground state wavefunction, (iii) fusion of non-Abelian anyons that results in an excited state with a \textit{single} anyon on the torus, and (iv) a non-trivial braiding sequence of three non-Abelions---also known as the ``Borromean rings''---where the absence of pairwise linking makes it invisible to Abelions (Fig.~\ref{fig_conceptual}b-c).

These experiments go beyond merely \emph{simulating} non-Abelian order and statistics.
As we elaborate in~\ref{methods_hardware}, the data obtained from prior benchmarking~\cite{moses2023race} is compatible with the fact that both the quantum operations as well as the dominant imperfections follow the 2D geometry to which the qubits are assigned. Thus, the ions \emph{are} entangled in precisely the same way as the low-energy states of Eq.~\eqref{eq_hamiltonian}, making them indistinguishable from states arising in the low-temperature limit of, e.g., a solid-state system governed by the same Hamiltonian. Ground state degeneracy then refers to the number of locally indistinguishable states, while anyons are local deformations which cannot be individually created by a local process.

The ability to create such exotic entangled states owes to recent developments in experiment and theory. First, while it is known that unitary circuits creating topological order have limited scalability due to their depth scaling with system size \cite{Bravyi06,Liu21} (even for long-range gates \cite{aharonov2018quantum}), it has long been known that combining a finite-depth circuit of unitaries with measurement and feed-forward can prepare \emph{Abelian} topological orders \cite{Raussendorf05,Bolt16,Piroli21,tantivasadakarn_hierarchy_2022}. On the experimental side, it has only recently become possible to implement such adaptive circuits, illustrated by a toric code preparation \cite{iqbal_topological_2023,Fossfeig23}. However, preparing \emph{non-Abelian} states posed a challenge, since a straightforward extension of this procedure creates unpaired non-Abelions which require linear-depth unitaries to remove \cite{Shi19}. This efficient preparation problem was recently solved for a special class of `solvable’ non-Abelian topological orders \cite{measureSPT,verresen2022efficiently,Bravyi22}, and was further simplified in the case of the quantum double~\cite{kitaev_fault-tolerant_2003} of the group $D_4$ ($\mathcal D(D_4)$) in Ref.~\onlinecite{tantivasadakarn_shortest_2022}. This will be the strategy adopted in the current work.

\begin{figure*}[!ht]
	\centering
    \includegraphics[width=1\textwidth]{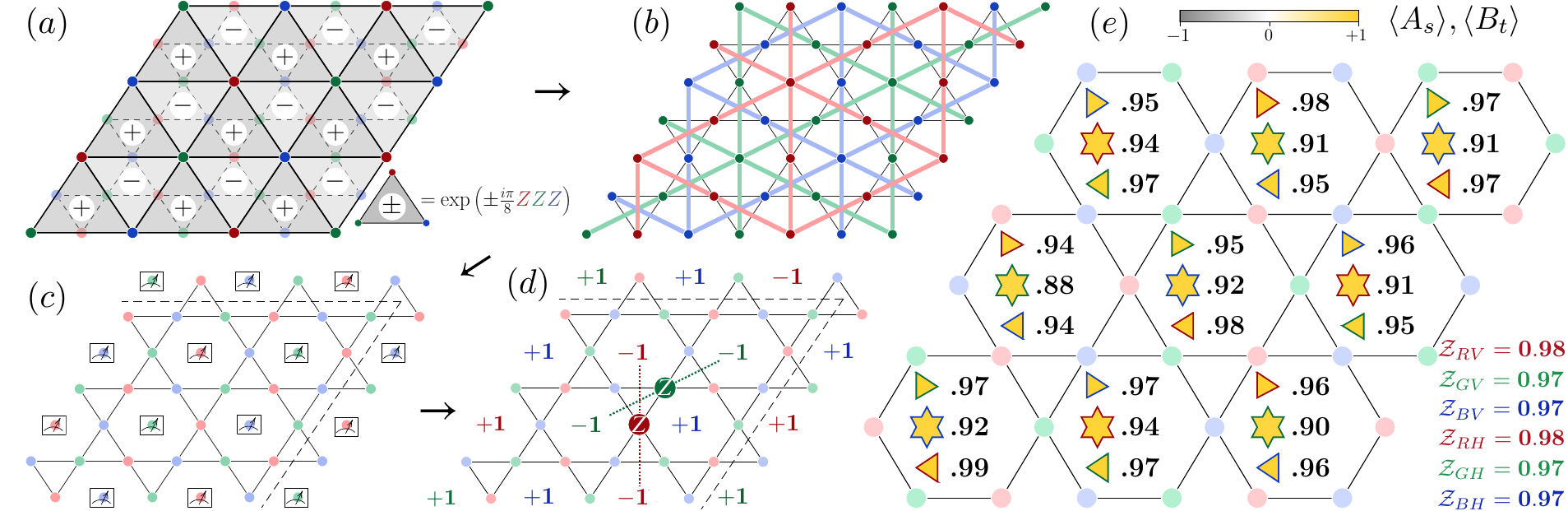}
	\caption{\textbf{Ground state preparation protocol and experimental data.} (a) Starting with a $|+\rangle^{\otimes N_P}$ product state on the plaquettes of the kagome lattice, we entangle all qubits via non-Clifford three-qubit $\exp(\pm i \pi/8 ZZZ)$ gates; the sign is $+1$ ($-1$) for up-pointing (down-pointing) triangles. (b) The plaquette qubits are subsequently entangled with qubits living on the vertices of the kagome lattice (see Eq.~\eqref{eq_gs_prep}) via $\textsc{CNOT}$ gates (colored lines). 
    (c) Measuring the plaquette qubits in the $X$-basis prepares a state with non-Abelian topological order on the kagome lattice, albeit with a random pattern of Abelian anyons where the 12-body term $A_s=-1$. (d) A feed-forward layer of conditional $Z$ gates is applied to pair up the Abelions, giving a state with $A_s = B_t = 1$. Dashed lines indicate the periodic boundary conditions. (e) After preparing the state, we experimentally measure the expectation values of star ($A_s$), triangle ($B_t$) and logical $\mathcal{Z}$ operators (see Fig.~\ref{fig_overview} for definitions). The measured energy density with respect to Eq.~\eqref{eq_hamiltonian} is $-0.946(4)$. The average (maximal) standard error on the mean of the star and triangle operators is 0.015 (0.022). The average (maximal) standard error on logical operators is 0.004 (0.005).}
 \label{fig_protocol}
\end{figure*}

\section{The model}
Our goal is to prepare and control eigenstates of a Hamiltonian of qubits arranged on a periodic kagome lattice giving rise to non-Abelian topological order \cite{yoshida_topological_2016}:
\begin{equation}
    H = - \sum_{s \in \{\davidsstar\}} A_s - \sum_{t \in \{ \triangleleft ,  	\triangleright\}} B_t.
\label{eq_hamiltonian}
\end{equation}
The 12-body star operator $A_s = \prod_{i=1}^6 CZ_{i,i+1} X^{\otimes 6}$ and 3-body triangle operator $B_t = Z^{\otimes 3}$ are shown in Fig.~\ref{fig_overview}. Each plaquette of the kagome lattice supports one $A_s$ and two $B_t$ terms. It will be convenient to choose a three-coloring of the kagome lattice, inducing a coloring of each star and triangle term cf. Fig.~\ref{fig_overview}; we will refer to this \red{red}-\green{green}-\blue{blue} coloring throughout the paper. While the Hamiltonian is known~\cite{yoshida_topological_2016}, we introduce simple expressions for the anyon and logical operators which are essential for its experimental realisation and verification.

Without the Controlled-$Z$ gates in $A_s$, model \eqref{eq_hamiltonian} constitutes three decoupled toric code Hamiltonians \cite{kitaev_fault-tolerant_2003} associated to (the bonds of) three triangular lattices in red, green and blue (Fig.~\ref{fig_protocol}b). The $CZ$ gates couple the toric codes, leading to the $A_s$ generally failing to commute with each other. Indeed, stabilizer Hamiltonians cannot support non-Abelian order~\cite{potter_symmetry_2016}. Nevertheless, in the subspace where $B_t = 1$, we have $[A_s, A_{s'}]$ = 0 and therefore all ground states fulfill $A_s = B_t = 1$. The violation of a star term $A_s=-1$ (triangle $B_t=-1$) signals the presence of an Abelion (a non-Abelion).

These states can be further distinguished by the value of logical operators on the torus. These act with $Z$ on all qubits of a given color in either the horizontal or vertical direction (Fig.~\ref{fig_overview}), and we will denote such operators by, e.g., \red{$\mathcal{Z}_{RH}$}, with subscripts $R$, $G$, $B$ denoting color and $H$, $V$ denoting the horizontal or vertical direction. We first focus on the case where all $\mathcal Z$-logicals are $+1$. Our last two sets of experiments explore the other logical sectors on the torus.

\begin{figure*}[!t]
	\centering
	\includegraphics[width=1\textwidth]{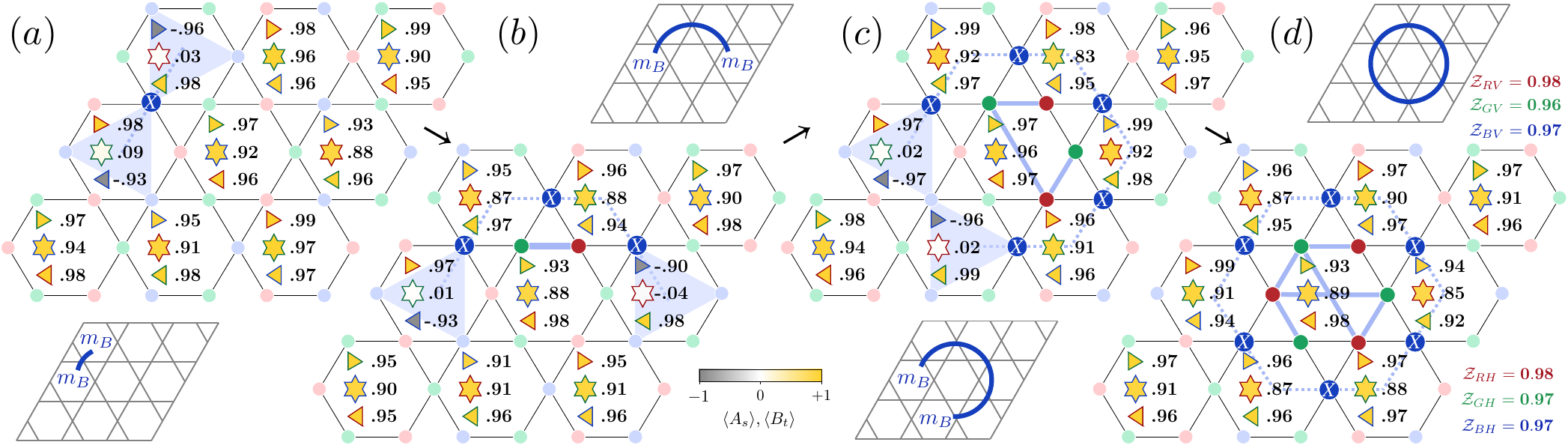}
	\caption{\textbf{Creating and fusing a non-Abelian anyon pair.} (a) A non-Abelion pair is created from the vacuum by an $X$-operator that toggles the two adjacent blue triangles $B_t$ (shaded). Exciting such a pair locally leads to $A_s = 0$. Physically, this indeterminate value of $A_s$ corresponds to the internal space of the non-Abelion. (b,c) Moving a non-Abelion requires a linear depth quantum circuit. To move a blue non-Abelion two stars over, we apply Pauli-$X$ operators on blue qubits (blue dashed) and $CZ$ operators between each red qubit and all preceding green qubits (thick blue). (d) Fusion of the blue non-Abelion pair in the identity channel brings the state back to $\ket{\psi_0}$. The average (maximum) standard error on the mean of star and triangle operators is 0.023 (0.068). The average (maximum) standard error on the mean of logical operators in (d) is 0.006 (0.008).
 }
 \label{fig_fuse_identity}
\end{figure*}

\section{Ground State Preparation}
\label{sec_gs_preparation}
Our first goal is to prepare the unique ground state $\ket{\psi_0}$ of Eq.~\eqref{eq_hamiltonian} with all logical $\mathcal{Z}$-operators equal to $+1$. We employ a constant-depth adaptive circuit on the vertices $V$ of the kagome lattice proposed in Ref.~\onlinecite{tantivasadakarn_shortest_2022}, based on:
\begin{equation}
\! \ket{\psi_0}_V = \bra{+}_P\!  \prod_{\braket{p,v}}\!\! \textsc{CNOT}_{p,v}\!\!\! \prod_{\langle p , \tilde p , \tilde{\tilde p}\rangle} \!\!\!e^{\pm \frac{i \pi}{8}Z_p Z_{\tilde p}Z_{\tilde{\tilde p}}}  \! \ket{+}_{P} \! \ket{0}_V
    \label{eq_gs_prep}
\end{equation}
where $\langle p , \tilde p , \vardbtilde{p} \rangle$ ranges over the triangular plaquettes of the ancillas $P$ shown in Fig.~\ref{fig_protocol}a. Unpacking Eq.~\eqref{eq_gs_prep} into a preparation protocol, we start with a product state $\ket{+}_{P}\ket{0}_V$ which satisfies $B_t =\mathcal Z = 1$. The protocol to also obtain $A_s = 1$ proceeds in three steps.

First, we entangle the ancillas $\ket{+}_P$ using $\exp(\pm i \pi/8 \, ZZZ)$-gates, giving a symmetry-protected topological (SPT) state with $\mathbb Z_2^3$ symmetry \cite{SPTreview,yoshida_topological_2016}. Subsequently, we entangle this SPT state to $\ket{0}_V$ by applying cluster state entanglers \cite{Briegel00} on the three sublattices shown in Fig.~\ref{fig_protocol}b (up to a Hadamard transformation on $V$). Lastly, we measure all ancillas in the $X$-basis. If the outcome is $+1$, then Eq.~\eqref{eq_gs_prep} shows we obtain the desired state; this has been interpreted as effectively gauging the $\mathbb Z_2^3$ symmetry of the SPT state \cite{measureSPT} which indeed gives $D_4$ non-Abelian topological order (see~\ref{appendix_anyons}). A $-1$ outcome signals the presence of an Abelian anyon on that star. These come in pairs for each color and applying appropriate $Z$-operators pair up the anyons at the end of the protocol (Fig.~\ref{fig_protocol}d). Using such active feed-forward, we achieve deterministic state preparation without post-selection!

We implemented the above protocol for a lattice containing $3 \times 3$ stars (27 qubits) on periodic boundary conditions. For this geometry, the ground state preparation naively requires $4 N_\text{stars} = 36$ qubits and $12 N_\text{stars} = 108$ two-qubit gates. Using circuit optimisation and qubit reuse techniques\footnote{In particular, certain ancillas used for mid-circuit measurements are reset and then reused; the high-fidelity reset implies this does not affect the effective 2D connectivity.}, we reduce these requirements to $30$ qubits and $78$ two-qubit gates (cf.~\ref{appendix_circuit_optimisation}). A barrier in the circuit ensures that the full state is prepared before any measurement of the star, triangle, or logical operators. In practice, noise may corrupt the state in such a way that an odd number of star excitations is measured in step three of the protocol. These errors are heralded and we choose to discard the corresponding states leading to an average discard rate of $15\%$ across all experiments reported (note that the scalability of the protocol is not affected~\cite{iqbal_topological_2023}). After preparing the state, we measure the star, triangle and logical operators by executing single-qubit measurements in three different settings, each corresponding to one of the sublattices.

The experimental results of this state preparation strategy are shown in Fig.~\ref{fig_protocol}e. The data implies a bound of $\sqrt[27]{\braket{\psi_0|\rho_\text{prepared}|\psi_0} }\geq  0.984(1)$ on the fidelity per site for the raw data. Note that such fidelity per site is a natural measure for large states, similar to an effective temperature. After correcting for readout error this increases to $0.990(1)$ giving a {\it global} fidelity $\geq 0.75(2)$ (see section \ref{app_fidelity_lower_bound} for a derivation). Remarkably, this fidelity per site is similar to a recent state-of-art constant-depth preparation of the much simpler toric code state \cite{iqbal_topological_2023}.

\begin{figure*}[!ht]
	\centering 
	\includegraphics[width=1.0\textwidth]{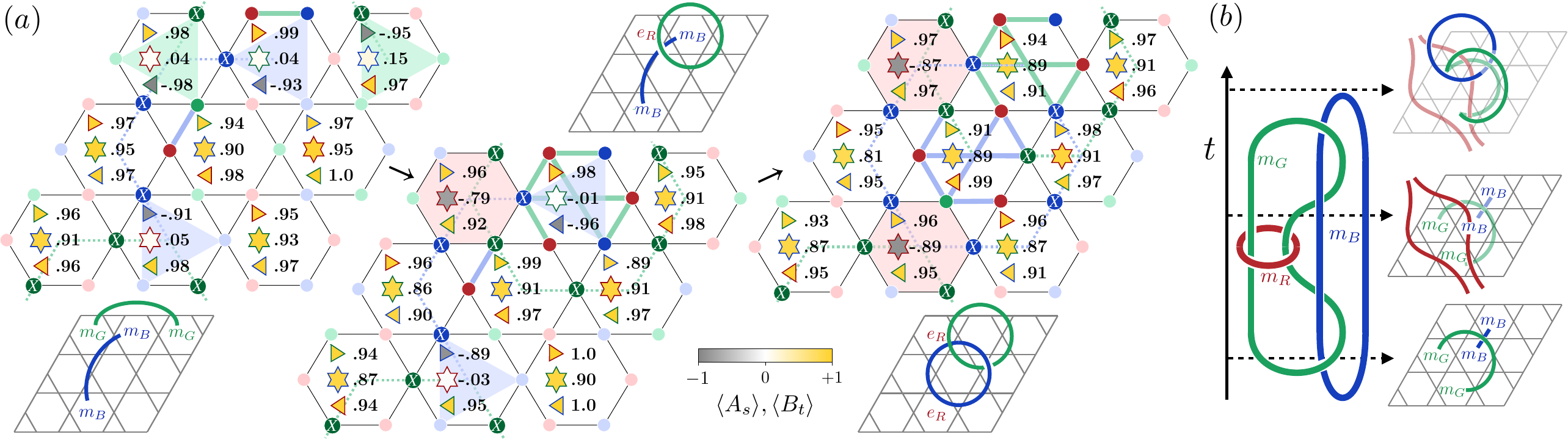}
    \caption{\textbf{Braiding non-Abelian anyons.}
    %The notation of the experimental snapshots follows Fig.~\ref{fig_protocol}e.
    (a) A pair of blue non-Abelian fluxes \blue{$m_B$} is created, indicated by excited triangle operators. Subsequently, one partner of a pair of green \green{$m_G$} is braided around one of the \blue{$m_B$} and annihilated, leaving behind an \red{$e_R$} (shaded) due to the fusion rules~(\ref{eq_fusion_rules}). The upper blue anyon is then brought back to its partner, creating another \red{$e_R$}. The average (maximum) standard error on the mean of star and triangle operators is 0.025 (0.069). (b) Borromean braiding in spacetime. Pairs of \blue{$m_B$}, \green{$m_G$} and \red{$m_R$} are created and braided in such a way that each pair is unlinked. Thumbnails show the operators applied at different points in time. The creation and movement of the the \blue{$m_B$} is controlled on an ancilla, allowing to extract the phase $1.02(2)\pi$, with a modulus $r = 0.80(2)$. For Abelian anyons the phase of the corresponding diagram would be 0 (see Fig.~\ref{fig_conceptual}).}
 \label{fig_fuse2red_borromean}
\end{figure*}

%\section{Non-Abelion Creation, Braiding and Annihilation}
\section{Braiding Non-Abelian Anyons}
\label{sec_braiding}
Having established a high-fidelity state preparation, we will now turn to the experimental exploration of the anyon content of the model. This phase of matter supports 22 anyons which are classified in section~\ref{appendix_anyons}. We focus on three which are single-color non-Abelian bosons, denoted \red{$m_R$}, \green{$m_G$} and \blue{$m_B$}, corresponding to $B_t=-1$.

While Abelian anyons can be moved using constant depth unitaries (and indeed we have employed such a layer of conditional $Z$-gates during the state preparation), separating a pair of non-Abelions necessarily requires a quantum circuit whose depth scales linearly with their distance \cite{Shi19,Liu21,Bravyi22}. In the present model, a pair of, e.g., \blue{$m_B$} on two blue triangles $t_i$ and $t_f$ (one pointing left, one pointing right) can be created with the operator
\begin{equation}
    \blue{X_{t_i}^{t_f}} = \prod_{\blue{b}} X_{\blue{b}} \prod_{\red{r} \green{g}} CZ_{\red{r} \green{g}}
    \label{eq_non_abelian_pair}
\end{equation}
where $\prod_{\red{r} \green{g}} CZ_{\red{r} \green{g}}$ is a linear-depth circuit of Controlled-$Z$'s. Its structure is such that, with the exception of the endpoints, we remain in the ground state (i.e., Eq.~\eqref{eq_non_abelian_pair} commutes with all $B_{t \notin \{ t_i,t_f\}}$, as well as all $A_s$ in the $B_t=1$ subspace). An example of such a string is shown in Fig.~\ref{fig_overview}, where it defines the logical $\mathcal{X}$-operator upon wrapping around a periodic direction. Fig.~\ref{fig_fuse_identity} contains more examples, where we show experimental results for the creation, movement and annihilation of a non-Abelion pair. The results are compatible with the string leaving the state invariant away from its endpoint. The state is observed to return to the initial ground state after annihilating the pair.

The above process can be interpreted in terms of the fusion rules of the model. Specifically, for the anyons considered, 
\begin{equation}
    \blue{m_B} \times \blue{m_B} = 1 + \red{e_R} + \green{e_G} + \red{e_R} \green{e_G}
    \label{eq_fusion_rules}
\end{equation}
where $\red{e_R}$ and $\green{e_G}$ are Abelian bosons corresponding to $A_s=-1$. Here, the formal sum on the right-hand side denotes the possible fusion outcomes when bringing two $\blue{m_B}$ together which captures the possible (superposition of) anyon types arising from fusing two anyons (see section \ref{appendix_anyons} for a more detailed description). The fusion of pairs of $\red{m_R}$ and $\green{m_G}$ is given by permutation of the colors in~(\ref{eq_fusion_rules}). The fusion channel of a single $\blue{m_B}$-pair created from the vacuum is necessarily the identity `1'.
In a topological quantum computer, the fusion channel is the degree of freedom that encodes the quantum information. The ability to change the internal state of the non-Abelions is a necessary requirement for topological quantum computation. Demonstrating this ability is what we turn to next.

Specifically, braiding $\green{m_G}$ around $\blue{m_B}$ toggles the fusion channel for both pairs from $1$ to $\red{e_R}$, essentially executing a Pauli-$X$-gate on the fusion channel space. The results of this braiding are reported in Fig.~\ref{fig_fuse2red_borromean}a. After the annihilation of both pairs, the state is found to be in the ground state, except at the points where $\blue{m_B}$ and $\green{m_G}$ were annihilated. There, negative values of the star operators indicate the presence of $\red{e_R}$-Abelions. The fusion product is not visible during the braiding process but only reveals itself upon annihilation of the anyons, evidencing the fact that information is stored non-locally in the entanglement of the wavefunction.

Even when there is no net linking between non-Abelian worldlines---such that we fuse back to the vacuum at the end---the non-Abelian fusion rules can lead to striking consequences. In particular, we can detect a three-anyon braiding which would be invisible for a purely Abelian state, even if it hosts non-Abelian defects. Concretely, pairs of \red{$m_R$}, \green{$m_G$} and \blue{$m_B$} are created, moved and annihilated in such a way to form ``Borromean rings" in spacetime (Fig.~\ref{fig_fuse2red_borromean}b)~\cite{wang_topological_2015,WangWenYau20,putrov_braiding_2017,kulkarni_topological_2021, chan_braiding_2018}. The defining property of this linking is that any pair of rings is unlinked such that the removal of any one of them renders the diagram topologically trivial. Thus, for any triplet of Abelian particles, such a braiding action must necessarily be trivial (cf. Fig.~\ref{fig_conceptual}). In stark contrast, in the present model, the ground state is theoretically predicted to pick up a minus sign under a Borromean braid. We have implemented a Hadamard test, confirming experimentally the braiding phase
\begin{equation*}
    \arg\braket{\psi_0| \green{X_g^{g'}} \red{X_r^r} \green{X_{g'}^g} \blue{X_{b'}^b} \green{X_{g'}^g} \red{X_r^r} \green{X_g^{g'}} \blue{X_b^{b'}} |\psi_0} = 1.02(2) \pi
\end{equation*}
(see Fig.~\ref{fig_fuse2red_borromean}b and Extended Data Figure \ref{EDF_borromean} for the microscopic definition of the operators). Removing the red or green action trivialises the action and we find phases of $0.00(2) \pi$ and $0.07(2) \pi$ for those braids, respectively.

\begin{figure*}[!ht]
	\centering
    \includegraphics[width=1\textwidth]{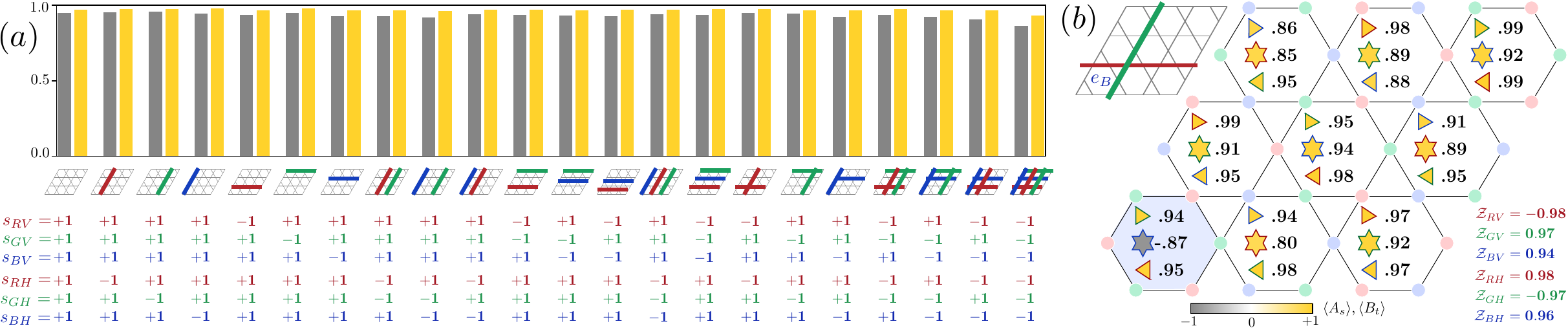}
    \caption{\textbf{Creation of all 22 ground states and a state with a single anyon on the torus.} (a) Logical $\mathcal{X}$-operators are applied by winding non-Abelian anyons around the torus. There are $2^6 = 64$ possible combinations, the 22 configurations shown here fulfill the constraint $A_s = 1$ for each color according to~(\ref{eq_constraint}). The bar plot shows experimentally measured values of the negative energy density (gray) and a logical pinning function (yellow) $\Pi := 1/6 \sum_{cd} s_{cd} \overline{\mathcal{Z}_{cd}}$ where $c$ and $d$ run over the three colors and two directions, respectively, $\overline{\mathcal{Z}}_{cd}$ is the corresponding logical $\mathcal{Z}$ operator averaged over translations and $s_{cd} = \pm 1$ is the sign of $\mathcal{Z}_{cd}$ in the target logical sector. This operator is equal to 1 if and only if the state is in the theoretically predicted logical sector. The average of the energy density (logical pinning function) across 22 ground states is $-0.93$ (0.97). The average (maximum) standard error on the mean of the energy densities and logical pinning functions is 0.007 (0.009) and 0.003 (0.004), respectively.
    (b) A red horizontal and green vertical $\mathcal X$ operator are applied to the ground state. The red non-Abelion flips the logical $\mathcal{Z}$ operator that toggles the fusion channel of the green pair. The state after this procedure is experimentally found to have a single blue Abelian excitation. The average (maximum) standard error on the mean of the energy and logicals is 0.017 (0.029) and 0.005 (0.007), respectively.
    }
 \label{fig_23gs}
\end{figure*}

\section{Logical sectors and single anyons}
\label{sec_noncontractible_loops}

Thus far, our experiments have created ground states and anyons in the logical $\mathcal{Z}=+1$-sector of the $D_4$ model. We can toggle between the logical-$\mathcal{Z}$ eigenstates by moving $m$ anyons around the torus. An example of such a logical-\blue{$\mathcal X_{BV}$} operator, which toggles \blue{$\mathcal Z_{BH}$}, is shown in Fig.~\ref{fig_overview}. As we now show, the non-Abelian nature of these anyons leads to striking consequences for these different sectors.

In principle, there are $2^6 = 64$ different assignments of $\pm 1$ to the six logical $\mathcal{Z}$-operators. However, in stark contrast to Abelian topological orders (like the toric code), there is a non-trivial constraint 
\begin{equation}
    %\prod_{\red{s \in \text{red \ding{65}}}}
    \prod_{\red{s \in \{\davidsstar\}}}
    \red{A_s} = (-1)^{\frac{1-\green{\mathcal{Z}_{GH}}}{2}\frac{1-\blue{\mathcal{Z}_{BV}}}{2}} \times (-1)^{\frac{1-\green{\mathcal{Z}_{GV}}}{2}\frac{1-\blue{\mathcal{Z}_{BH}}}{2}}
    \label{eq_constraint}
\end{equation}
between the star and logical operators in the $B_t = 1$ subspace and analogous relations hold for the other colours (see~\ref{appendix_uniqueness_and_gs_degeneracy} for a derivation). To keep the right-hand side of Eq.~\eqref{eq_constraint} positive, we can apply any combination of horizontal $\mathcal{X}_{\lambda H}$-logicals, \emph{or} vertical $\mathcal{X}_{\lambda V}$-logicals, \emph{or} $\mathcal{X}_{\lambda H}\mathcal{X}_{\lambda V}$-logicals. This leads to a ground state degeneracy of $1+(2^3-1)\times 3=22$. It can be shown that there are no other ground states (see~\ref{appendix_uniqueness_and_gs_degeneracy}). We have experimentally created all 22 states by applying the above logical-$\mathcal X$ operators to $\ket{\psi_0}$. The main results are shown in Fig.~\ref{fig_23gs}a, with a full set of expectation values reported in Extended Data Tables~\ref{table:22_gss_a} and~\ref{table:22_gss_b}. We have thus demonstrated the control necessary to experimentally access the full ground state subspace.

The existence of a ground state degeneracy that is not a perfect square for a state that admits a gapped boundary is a striking consequence of the non-Abelian order. This is further emphasized by considering the fate of logical sectors where the right-hand side of Eq.~\eqref{eq_constraint} is negative. Apparently, such states host an \emph{odd} number of Abelian anyons. We experimentally confirm this curious prediction by applying \green{$\mathcal X_{GV}$}\red{$\mathcal X_{RH}$} to $\ket{\psi_0}$. Strikingly, we measure a \emph{single} anyon \blue{$e_B$} in the resulting state (Fig.~\ref{fig_23gs}b)! More generally, from the $2^6$ sectors, 42 contain such unpaired anyons, leaving 22 ground states.

\section{Outlook}
In this work we have demonstrated the creation of ground states with non-Abelian topological order as well as the controlled braiding of non-Abelion anyons and tunneling between logical states on a trapped-ion quantum processor. Our key finding is that non-Abelian topological orders can experimentally be prepared with high fidelities on par with Abelian states like the surface code. Non-Abelian states are among the most intricately entangled quantum states theoretically known to exist, and carry promises for new types of quantum information processing. Their realization evidences the rapid development of quantum devices and opens up several new questions.

Unlike ground states of Abelian topological order, logical states in the $D_4$-model can only be toggled via a linear-depth circuit. Whether this property leads to a greater tolerance against bit-flip noise is an interesting question to explore. In the runtime of the current experiment, the state fidelity and qubit lifetime did not require any stabilization to achieve our results. In future works, rather than utilizing energy conservation to protect our state as one would in Hamiltonian set-ups,  we envisage stabilizing it by repeatedly measuring the terms in Eq.~\eqref{eq_hamiltonian} and pairing up the anyons using the above string operators. A fault-tolerant threshold for $D_4$ topological order has been  been argued for \cite{Dauphinais2017}, and a deeper exploration of these issues is left to future work.

More broadly, it would be fascinating to explore other novel types of entangled phases of matter. Certain `solvable' non-Abelian states can be accessed by utilizing multiple rounds of measurement \cite{measureSPT,verresen2022efficiently,tantivasadakarn_hierarchy_2022,Bravyi22}, whereas others can be obtained from log-depth adaptive circuits \cite{Hsieh22}. Remarkably, finite-depth adaptive circuits have even been proposed to create states with algebraic correlations \cite{Zhu22,Lee22,Lu23}.

Finally, it is not known whether the gate set implemented by measurement and braiding of anyons with $D_4$ order can be made universal. However, universal $S_3$ states~\cite{mochon_anyon_2004, cui_universal_2015} can be created by adding a single layer of feed-forward \cite{tantivasadakarn_hierarchy_2022}. Our work thus points a way toward demonstrating the first universal gate set from braiding non-Abelions and measurement.

\bibliography{references2}

%\small

%\clearpage
\onecolumngrid
\newpage

\appendix

\section{Methods}

\subsection{Important aspects of the physical realisation}
\label{methods_hardware}

In this section we argue why the entanglement of the state as well as the leading sources of error follow the two-dimensional kagome geometry introduced in the main text. For a full description and characterisation of the device used in the experiment, see Ref.~\cite{moses2023race}. The ions are stored in a 1D ring, but their spatial ordering is dynamically and repeatedly reconfigured during execution of a quantum circuit in order to bring arbitrary pairs together for applying laser-driven two-qubit gates. These gates are the dominant source of error in the device, and operate with an average fidelity of about 99.82\%. Despite the storage of ions in a 1D ring, in the state preparation and braiding protocols they are immutably assigned to the vertices of the 2D lattice defined by the Hamiltonian in Eq.~\eqref{eq_hamiltonian}, and all gates are local in that 2D geometry. Moreover, the extremely long coherence times of trapped ion qubits and low cross-talk afforded by the quantum charge coupled device architecture (and systematically characterized in Ref. \cite{moses2023race}) ensure that the dominant noise processes are local and uncorrelated errors attached to each two-qubit gate. Thus even the dominant \emph{imperfections} in our creation of these states respect the 2D geometry defined by Eq.~\eqref{eq_hamiltonian}.

\subsection{Proof of~(\ref{eq_constraint}) and colour algebra}
\begin{figure}[!ht]
	\centering 
	\includegraphics[width=1.0\textwidth]{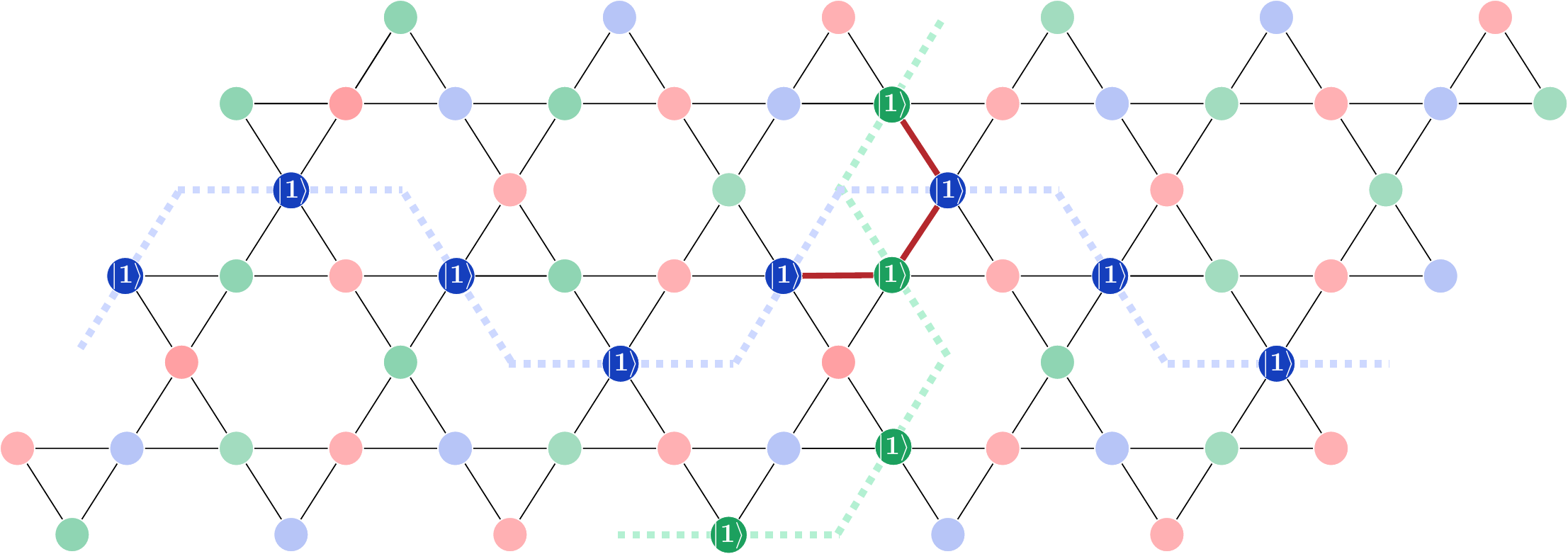}
    \caption{\textbf{Evaluating the product $\prod_{\red{s \in \text{red \ding{65}}}} \red{A_s}$ for a computational basis state.} 
    Shown is a computational basis state with an odd number of blue strings in the horizontal and green strings in the vertical direction on which $(1-\green{\mathcal{Z}_{GH}})/2 \times (1-\blue{\mathcal{Z}_{BV}})/2 = +1$.
    There is an odd number of red stars (one) where these strings cross. By considering the $2^6$ possible bitstrings around a hexagon, one can verify that $\prod CZ = -1$ around that hexagon if and only if the corresponding star is such a crossing star. Here, the only three $CZ$ in the product $\prod_{\red{s \in \text{red \ding{65}}}} \red{A_s}$ which evaluate to $-1$ are marked with solid red lines.}
    \label{EDF_color_algebra}
\end{figure}

Here we prove the relationship~(\ref{eq_constraint}) between the star and logical operators, reproduced here for convenience:
\begin{equation}
    \prod_{\red{s \in \text{red \ding{65}}}} \red{A_s} = (-1)^{\frac{1-\green{\mathcal{Z}_{GH}}}{2}\frac{1-\blue{\mathcal{Z}_{BV}}}{2}} \times (-1)^{\frac{1-\green{\mathcal{Z}_{GV}}}{2}\frac{1-\blue{\mathcal{Z}_{BH}}}{2}}
\end{equation}
For ease of notation, one of the three colours is singled out but it is understood that equivalent statements hold for all permutations of colours and directions. By linearity, it suffices to show that the equation holds for all computational basis states. Since we are working in the $B_t = 1$ subspace, strings of $\ket{1}$s of a given color must form closed loops on the honeycomb superlattices of that color. These loops can either be contractible or wrap around the torus. Therefore e.g., $\green{Z_{GH}}$ acts on computational basis states by counting the parity of strings of $\ket{1}$s on green qubits wrapping around the torus in the vertical direction. The operator $(1-\green{\mathcal{Z}_{GH}})/2 \times (1-\blue{\mathcal{Z}_{BV}})/2$ projects into the space of computational basis states that have an odd number of blue strings in the horizontal and green strings in the vertical direction. In this space, there must be an odd number of stars where these strings cross (and these stars must necessarily be red). On the other hand, the product of $CZ$s within a red star is $-1$ if and only if the star is such a crossing star, as can be verified by considering rotations of Extended Data Figure~\ref{EDF_color_algebra}. Going through the same argument for the other colours and torus directions concludes the proof.

Note that~(\ref{eq_constraint}) implies the ``color algebra" (that we note here for completeness)
\begin{align}
    \prod_{\red{s \in \text{red \ding{65}}}} \red{A_s} \blue{\mathcal{X}_{BH}} &= \green{\mathcal{Z}_{GH}} \blue{\mathcal{X}_{BH}} \prod_{\red{s \in \text{red \ding{65}}}} \red{A_s}
\end{align}
which follows by using $\blue{\mathcal{Z}_{BH} \mathcal{X}_{BV}} = - \blue{\mathcal{X}_{BV} \mathcal{Z}_{BH}}$.

\subsection{Uniqueness of the $\mathcal{Z}=1$ state and Ground State Degeneracy}
\label{appendix_uniqueness_and_gs_degeneracy}
Here we show the uniqueness of the ground state of~(\ref{eq_hamiltonian}) in a given logical sector, and show that there are exactly 22 logical sectors that contain ground states. These proofs hold for tori of arbitrary sizes, not just the $3 \times 3$-torus implemented in the experiment. All of the statements in this section hold within the $B_t = 1$-subspace.

We start by fixing a logical sector through the specification of a set of $z_{cd} = \pm 1$. Denote the set of kagome stars on the lattice by $S$ (in the experiment $|S|=9$). Select one red (\red{$s_R$}), one green ($\green{s_G}$) and one blue star ($\blue{s_B}$). A counting argument reveals that the subspace defined by
\begin{align}
    \left\{ \mathcal{Z}_{cd} = z_{cd}, B_t = 1, A_s = 1 \bigg| c \in \{\red{R}, \green{G}, \blue{B}\}, d\ \in \{H, V\}, \forall t, s \in S \, \backslash\,  \{\red{s_R},\green{s_G},\blue{s_B}\}   \right\}
\end{align}
has dimension one. There are 6 logical $z_{cd} \mathcal{Z}_{cd}$ stabilisers and $2|S|$ triangular stabilisers, $2|S|-3$ of which are independent due to $\prod_t B_t = 1$ individually for each of the colors, leading to $2|S|+3$ independent stabilisers that are diagonal in the computational basis. The independence of these diagonal stabilisers follows from the same argument as in the toric code (products of $Z$-plaquettes wrap around the torus an even number of times). The $A_s$ operators are independent from the logical and triangle operators since any product that does not involve all $A_s$ of a given color (including \red{$s_R$}, \green{$s_G$} or \blue{$s_B$}) contains off-diagonal terms. Finally, the $A_s$ are mutually independent. To see this, pick any state in the correct logical sector with $A_s = \pm 1$. Then, this state can be transformed into a state with any other pattern of $A_s = \pm 1$ by connecting the plaquettes which are to be toggled to \red{$s_R$}, \green{$s_G$} or \blue{$s_B$}, (depending on their color) using strings of $Z$ operators (similar to the cleanup step of the protocol in Fig.~\ref{fig_protocol}). These operations do not change the $B_t$ and logical operators. Since we have found $3|S|$ independent stabilizers on $3|S|$ qubits (and they commute in the $B_t = +1$ space) the common +1-eigenspace is exactly one-dimensional.

The state $\ket{z_{cd}}$ in this one-dimensional space can now either be a ground state or a state with energy 2, 4 or 6 above the ground state, depending on the value of \red{$A_{s_R}$}, \green{$A_{s_G}$} and \blue{$A_{s_B}$}. Their values now simply follow from equation~(\ref{eq_constraint}) since in the space defined above e.g.,
\begin{align}
    \red{A_{s_R}} &= \prod_{\red{s \in \text{red \ding{65}}}} \red{A_s} \nonumber \\
    &= (-1)^{\frac{1-\green{\mathcal{Z}_{GH}}}{2}\frac{1-\blue{\mathcal{Z}_{BV}}}{2}} \times (-1)^{\frac{1-\green{\mathcal{Z}_{GV}}}{2}\frac{1-\blue{\mathcal{Z}_{BH}}}{2}}
\end{align}
Therefore, $\prod_{\red{s \in \text{red \ding{65}}}} \red{A_s} = \prod_{\green{s \in \text{green \ding{65}}}} \green{A_s} = \prod_{\blue{s \in \text{blue \ding{65}}}} \blue{A_s} = +1$ in the logical sectors with an even number of colour-pair crossings, which, for completeness, is the set
\begin{align*}
    (\red{\mathcal{Z}_{RH}}, \green{\mathcal{Z}_{GH}}, \blue{\mathcal{Z}_{BH}}, \red{\mathcal{Z}_{RV}}, \green{\mathcal{Z}_{GV}}, \blue{\mathcal{Z}_{BV}}) &\in \{ \red{0}\green{0}\blue{0}\red{0}\green{0}\blue{0}, \\
    &\quad \,\,\,\,
    \red{0}\green{0}\blue{0}\red{0}\green{0}\blue{1},
    \red{0}\green{0}\blue{0}\red{0}\green{1}\blue{0},
    \red{0}\green{0}\blue{0}\red{0}\green{1}\blue{1},
    \red{0}\green{0}\blue{0}\red{1}\green{0}\blue{0},
    \red{0}\green{0}\blue{0}\red{1}\green{0}\blue{1},
    \red{0}\green{0}\blue{0}\red{1}\green{1}\blue{0},
    \red{0}\green{0}\blue{0}\red{1}\green{1}\blue{1},\\
    &\quad \,\,\,\,
    \red{0}\green{0}\blue{1}\red{0}\green{0}\blue{0},
    \red{0}\green{1}\blue{0}\red{0}\green{0}\blue{0},
    \red{0}\green{1}\blue{1}\red{0}\green{0}\blue{0},
    \red{1}\green{0}\blue{0}\red{0}\green{0}\blue{0},
    \red{1}\green{0}\blue{1}\red{0}\green{0}\blue{0},
    \red{1}\green{1}\blue{0}\red{0}\green{0}\blue{0},
     \red{1}\green{1}\blue{1}\red{0}\green{0}\blue{0},
    \\
    &\quad \,\,\,\,
    \red{0}\green{0}\blue{1}\red{0}\green{0}\blue{1},
    \red{0}\green{1}\blue{0}\red{0}\green{1}\blue{0},
    \red{0}\green{1}\blue{1}\red{0}\green{1}\blue{1},
    \red{1}\green{0}\blue{0}\red{1}\green{0}\blue{0},
    \red{1}\green{0}\blue{1}\red{1}\green{0}\blue{1},
    \red{1}\green{1}\blue{0}\red{1}\green{1}\blue{0},
    \red{1}\green{1}\blue{1}\red{1}\green{1}\blue{1}  \} .
\end{align*}
where, for readability, we have labeled such bit strings with the values of the projectors $(1+\mathcal{Z})/2$ instead of $\mathcal{Z}$ (i.e., 0 or 1 instead of $\pm 1$).

In the absence of an analytic proof, the measurement-based protocol would also allow for an experimental detection of the ground state degeneracy. To this end, a random state is first prepared on all data qubits. In a second and third step the projections $\prod_t (1+B_t)/2$ and $\prod_s (1+A_s)/2$ are applied (in that order) via coupling to and measurement of ancillary qubits. Finally the logical $\mathcal{Z}$-operators are measured. Since, on average, random states have the same overlap with all ground state sectors, we expect each of the 22 ``allowed" bitstrings to appear with probability $\sim(1-\text{noise})/22$, while the 42 bitstrings ``forbidden" bitstrings appear with much lower probability $\sim \text{noise}/42$. To produce initial random states, approximate circuits may be sufficient, for example random Clifford circuits, or even random one-qubit unitaries, which have been shown to lead to reasonable result in the context of random measurements for the determination of entanglement entropies.

\subsection{Fidelity Lower Bound}
\label{app_fidelity_lower_bound}
Here, we show how to bound the fidelity per site from the experimentally measured correlation functions. Specifically we compute a lower bound on the the fidelity of the prepared state $\rho$ with respect to the unique state that is the $+1$ eigenstate of the star, triangle and logical operators $A_s$, $B_t$ and $\mathcal{Z}_{cd}$ (Fig.~\ref{fig_protocol}). We introduce the projectors (not to be confused with the braiding operators in the Borromean interferometry experiment)
%\begin{widetext}
\begin{align}
\begin{split}
    \red{R} &= \,\,\, \prod_{\red{s \in \text{red \ding{65}}}} \,\,\red{\frac{1+A_s}{2}}
    \prod_{\green{t \in \text{green} \triangleright}} \green{\frac{1+B_t}{2}} \prod_{\blue{t \in \text{blue} \triangleright}} \blue{\frac{1+B_t}{2}}
    \,\,\frac{1+\green{\mathcal{Z}_{GH}}}{2} \frac{1+\green{\mathcal{Z}_{GV}}}{2} \\
    \green{G} &= \prod_{\green{s \in \text{green \ding{65}}}} \green{\frac{1+A_s}{2}}
    \, \prod_{\blue{t \in \text{blue} \triangleright}} \, \blue{\frac{1+B_t}{2}} \,\, \prod_{\red{t \in \text{red} \triangleright}} \red{\frac{1+B_t}{2}}
    \,\,\frac{1+\blue{\mathcal{Z}_{BH}}}{2} \frac{1+\blue{\mathcal{Z}_{BV}}}{2} \\
    \blue{B} &= \, \prod_{\blue{s \in \text{blue \ding{65}}}} \,\, \blue{\frac{1+A_s}{2}}
    \,\,\prod_{\red{t \in \text{red} \triangleright}} \red{\frac{1+B_t}{2}} \, \prod_{\green{t \in \text{green} \triangleright}} \green{\frac{1+B_t}{2}}
    \,\, \frac{1+\red{\mathcal{Z}_{RH}}}{2} \frac{1+\red{\mathcal{Z}_{RV}}}{2}.
\end{split}
\end{align}
%\end{widetext}
Here, the products run over all stars of a given color except one (which can be chosen arbitrarily) and all triangles inscribed in the stars of that color, where across all three operators one arbitrary triangle can be excluded from each color. It follows that
\begin{align}
    [\red{R}, \green{G}] &= [\green{G}, \blue{B}] = [\blue{B},\red{R}] = 0 \\
    \ket{\psi_0} \bra{\psi_0} &= \red{R} \green{G} \blue{B}
\end{align}
We now seek to bound $\braket{\psi_0| \rho| \psi_0}$ by the measured expectation values $\red{\braket{R}} = \text{Tr}\rho \red{R}$, $\green{\braket{G}}$ and $\blue{\braket{B}}$. Since $\red{R}$, $\green{G}$ and $\blue{B}$ are commuting projectors, there exists a common orthonormal eigenbasis $\ket{\psi^{\red{r}\green{g}
\blue{b}}_k}$ with $\red{r},\green{g},\blue{b} \in \{0,1\}$ that fulfills 
\begin{align}
\label{eq_methods_RGB}
\begin{split}
    \red{R} \ket{\psi^{\red{r}\green{g}\blue{b}}_k} &= \red{r} \ket{\psi^{\red{r}\green{g}\blue{b}}_k} \\
    \green{G} \ket{\psi^{\red{r}\green{g}\blue{b}}_k} &= \green{g} \ket{\psi^{\red{r}\green{g}\blue{b}}_k} \\
    \blue{B} \ket{\psi^{\red{r}\green{g}\blue{b}}_k} &= \blue{b} \ket{\psi^{\red{r}\green{g}\blue{b}}_k}
\end{split}
\end{align}
and $k$ is a degeneracy label. Expanding the prepared state in this basis
\begin{align}
    \rho = \sum p^{\red{r}\green{g}\blue{b}k}_{\red{r'}\green{g'}\blue{b'}k'} \ket{\psi^{\red{r}\green{g}
\blue{b}}_k} \bra{\psi^{\red{r'}\green{g'}
\blue{b'}}_{k'}}
\end{align}
defines the coefficients $p^{\red{r}\green{g}\blue{b}k}_{\red{r'}\green{g'}\blue{b'}k'}$. Let us also introduce shorthand notation
\begin{align}
    P^{\red{r}\green{g}\blue{b}} &:= \sum_k p^{\red{r}\green{g}\blue{b}k}_{\red{r}\green{g}\blue{b}k}
\end{align}
for the sum of the diagonal elements of $\rho$ in a given sector in the basis chosen above. In this notation, the ground state fidelity is simply $\braket{\psi_0 | \rho |\psi_0} =  P^{\red{1}\green{1}\blue{1}}$. We then have

\begin{subequations}
\label{eq_fidelity_bound_all}
\begin{align}
    \red{\braket{R}} &= P^{\red{1} \green{1} \blue{1}} + P^{\red{1} \green{1} \blue{0}} + P^{\red{1} \green{0} \blue{1}} + P^{\red{1} \green{0} \blue{0}} \label{eq_fidelity_bound_a} \\
    \green{\braket{G}} &= P^{\red{1} \green{1} \blue{1}} + P^{\red{1} \green{1} \blue{0}} + P^{\red{0} \green{1} \blue{1}} + P^{\red{0} \green{1} \blue{0}} \label{eq_fidelity_bound_b} \\
    \blue{\braket{B}} &= P^{\red{1} \green{1} \blue{1}} + P^{\red{0} \green{1} \blue{1}} + P^{\red{1} \green{0} \blue{1}} + P^{\red{0} \green{0} \blue{1}} \label{eq_fidelity_bound_c} \\
    1 &= P^{\red{0} \green{0} \blue{0}} + P^{\red{0} \green{0} \blue{1}} + P^{\red{0} \green{1} \blue{0}} + P^{\red{0} \green{1} \blue{1}} + P^{\red{1} \green{0} \blue{0}} + P^{\red{1} \green{0} \blue{1}} + P^{\red{1} \green{1} \blue{0}} + P^{\red{1} \green{1} \blue{1}}
    \label{eq_fidelity_bound_d} 
\end{align}
\end{subequations}
By considering~(\ref{eq_fidelity_bound_a})+(\ref{eq_fidelity_bound_b})+(\ref{eq_fidelity_bound_c})-2(\ref{eq_fidelity_bound_d}), we see that $P^{\red{1} \green{1} \blue{1}}$ is minimized if $P^{\red{1} \green{0} \blue{0}} = P^{\red{0} \green{1} \blue{0}} = P^{\red{0} \green{0} \blue{1}} = P^{\red{0} \green{0} \blue{0}} = 0$. In this case 
\begin{align}
P^{\red{1} \green{1} \blue{1}} &= \red{\braket{R}} + \green{\braket{G}} + \blue{\braket{B}} - 2
\end{align}
and this is the lower bound for the fidelity. We have measured
\begin{align}
\begin{split}
    \red{\braket{R}} &= 0.90(1) \\
    \green{\braket{G}} &= 0.85(1) \\
    \blue{\braket{B}} &= 0.89(1)
\end{split}
\end{align}

\noindent leading to a lower bound on the global fidelity of the prepared state
\begin{align}
    \braket{\psi_0 | \rho | \psi_0} \geq 0.65(2).
\end{align}
The fidelity per qubit is
\begin{align}
    \sqrt[27]{\braket{\psi_0 | \rho | \psi_0}} & \geq 0.984(1).
\end{align}

\begin{figure*}[!ht]
	\centering
	\includegraphics[width=0.95\textwidth]{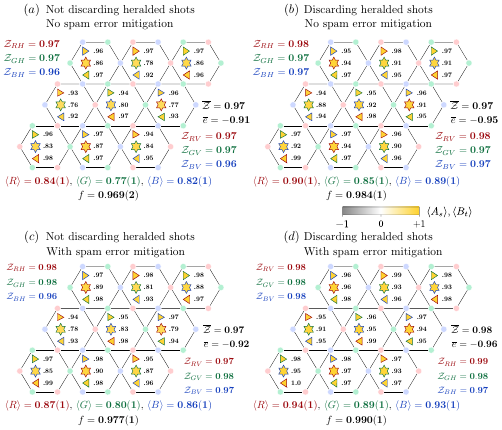}
	\caption{\textbf{Effect of discarding heralded errors and measurement errors.} ``Heralded shots" refers to the runs of the circuit where the measurement step in the ground state preparation protocol (Fig.~\ref{fig_protocol}c) reveals an odd number of Abelions due to noise. The spam error mitigation is described in section~\ref{app_fidelity_lower_bound}. The average (maximal) standard error on the mean of the star and triangle operators in (a-d) is 0.014 (0.027). The average (maximal) standard error on logical operators in (a-d) is 0.007 (0.01).
 $f$ denotes the lower bound on the resulting fidelity per qubit, calculated from $\red{R}$, $\green{G}$ and $\blue{B}$ as defined in~(\ref{eq_methods_RGB}).
 All main text figures correspond to the setting chosen in (b). 
 }
	\label{fig_spam_comparison}
\end{figure*}

These measurements do not take into readout errors. If one wants to assess the quality of the prepared state rather than the combined quality of state preparation and measurements, one must take into account the expectation values of \red{$R$}, \green{$G$} and \blue{$B$} after measurement error mitigation. Based on prior characterisation of the measurement error transition matrix in the device, $p(\text{measure } 0 | \text{qubit is } 1) = \num{2.37e-3}$, $p(\text{measure } 1 | \text{qubit is } 0) =  \num{0.82e-3}$, we have computed the corrected values by writing the raw probability distribution as a matrix product state of bond dimension $n_\mathrm{shots}$ and applying the transition matrix inverse on each site (see Extended Data Figure~\ref{fig_spam_comparison} for a detailed comparison). In principle, this correction accounts for both state preparation and measurement (SPAM) error. In practice, however, measurement errors dominate state preparation errors in the device.
We find
\begin{align}
\begin{split}
    \red{\braket{R}}_{\text{SPAM error mitigated}} &= 0.94(1) \\
    \green{\braket{G}}_{\text{SPAM error mitigated}} &= 0.89(1) \\
    \blue{\braket{B}}_{\text{SPAM error mitigated}} &= 0.93(1)
\end{split}
\end{align}
We infer that the prepared state actually has a global fidelity of
\begin{align}
    \braket{\psi_0 | \rho | \psi_0}_{\text{SPAM error mitigated}} \geq 0.75(2)
\end{align}
and a fidelity per qubit
\begin{align}
    \sqrt[27]{\braket{\psi_0 | \rho | \psi_0}}_{\text{SPAM error mitigated}} & \geq 0.990(1).
\end{align}

Using equation~(\ref{eq_fidelity_bound_all}), we can also bound the fidelity from above by noticing that
\begin{align}
    P^{\red{1} \green{1} \blue{1}} &\leq \min\Big\{ \red{\braket{R}},\green{\braket{G}},\blue{\braket{B}}\Big\}
\end{align}
We thus find
\begin{align}
    \braket{\psi_0 | \rho | \psi_0} &\in [0.65(2), 0.85(2)] \\
    \sqrt[27]{\braket{\psi_0 | \rho | \psi_0}} &\in [0.984(1),0.9940(7)]
\end{align}
without SPAM error mitigation and 
\begin{align}
    \braket{\psi_0 | \rho | \psi_0}_{\text{SPAM error mitigated}} &\in [0.75(2),0.88(1)] \\
    \sqrt[27]{\braket{\psi_0 | \rho | \psi_0}}_{\text{SPAM error mitigated}} &\in [0.990(1), 0.9955(6)]
\end{align}
with SPAM error mitigation. This is compatible with the fact that the state preparation protocol uses  $3 \, \nicefrac{1}{3}$ two-qubit gates per qubit with fidelity $99.82\%$ ($99.82\%^{3.3} \approx 99.4\%$) and the fact that gate errors typically dominate errors arising from dephasing and measurement cross-talk in the trap.

\subsection{Classification of the anyons}\label{appendix_anyons}

In the main text, the Hamiltonian corresponds to a gauged $\mathbb Z_2^3$ Symmetry-Protected Topological state~\cite{yoshida_topological_2016} and therefore is in the same family of models as the twisted quantum double $\mathcal D^\alpha(\mathbb Z_2^3)$~\cite{dijkgraaf_topological_1990,dijkgraaf_quasi_1991,hu_twisted_2013}. This model exhibits the same topological order as that of the quantum double of $\mathcal D(D_4)$~\cite{propitius_topological_1995,lootens_mapping_2022}. We elect to present the anyon content based on the twisted quantum double. A mapping relating the two conventions can be found in Extended Data Table~\ref{tab:D4anyons}.

First, recall that in the usual toric code, anyons are generated by an Abelian charge $e$ and an Abelian flux $m$ where $e^2 = m^2 =1$. Both anyons are bosons, but they have -1 mutual statistics. The bound state $em$ is an Abelian fermion. For three copies, all anyons of the toric code are generated by $e_C$ and $m_C$ where $C \in \{ R,G,B\}$ is a color index for each copy.

The 22 anyons of the twisted quantum double $\mathcal D^\alpha(\ZZ_2^3)$ can be labeled similar to that of three copies of the toric code. Instead, all fluxes are non-Abelian.

\begin{enumerate}
\item Eight Abelian bosons generated by $e_R$, $e_G$, $e_B$. Because they are Abelian, they obey the usual fusion rules i.e. $e_R \times e_G = e_{RG}$.
\item Three non-Abelian bosons $m_{R}, m_{G},m_{B}$. They braid with the corresponding charge of the same color with a $-1$ phase. I.e., $m_R$ braids non-trivially with $e_R$, but trivially with $e_G$ and $e_B$.
\item Three non-Abelian fermions $f_{R} = m_{R} \times e_R$. 
\item Three non-Abelian bosons $m_{RG},m_{GB}, m_{RB}$. One can interpret these as fluxes that respond to two colors.  That is $m_{RG}$ braids with both $e_R$ and $e_G$, but braids trivially with $e_B$.
\item Three non-Abelian fermions $f_{RG} = m_{RG} \times e_R = m_{RG} \times e_G$
\item A non-Abelian semion $s_{RGB}$, which is a flux that responds to three colors. That is, it braids with $e_R$, $e_G$, and $e_B$.
\item A non-Abelian antisemion $\bar s_{RGB} = s_{RBG} \times e_{RGB}$
\end{enumerate}

We summarize all the remaining fusion rules (up to permutation of colors and fusion of Abelian charges) in Extended Data Table~\ref{tab:D4fusion}. For example, by permuting colors, one can infer the fusion rule $m_G \times m_G = (1+e_R)(1+e_B)$ by permuting colors, and by the fusion rule $f_R \times m_G = e_R \times (m_R \times m_G) = e_R \times (m_{RG}+f_{RG}) = f_{RG} + m_{RG}$.

Next, we describe qualitatively why braiding $m_G$ around $m_B$ toggles the fusion outcome of each non-Abelian flux pair to $e_R$. The twisted quantum double realizes a gauged Symmetry-Protected Topological (SPT) phase given by the cocycle $\alpha$. Such SPT phase can be realized using a decorated domain wall construction~\cite{chen_symmetry-protected_2014,yoshida_topological_2016}. Specifically, the SPT phase can be realized by a superposition of blue domain walls decorated by 1D cluster states, which itself forms a 1D SPT state under the red and green symmetries. After gauging the symmetry, the blue domain wall is allowed to end, forming the deconfined excitation $m_B$. Furthermore, because the domain wall was decorated with a cluster state, its end point $m_B$ now carries a two-dimensional projective representation inherited from the end point of the cluster state, and therefore becomes a non-Abelian excitation due to this degeneracy.

The 1D cluster state protected by the red and green symmetries has the property that the ground state under antiperiodic boundary conditions of the green symmetry is odd under the red symmetry. Such a boundary condition can be enforced by introducing a single domain wall. Therefore, by braiding $m_G$ around $m_B$, a green domain wall cuts through the string operator of $m_B$. Inheriting the property from the cluster state, the entire $m_B$ string is now charged under the red gauge symmetry, which implies that fusing back the $m_B$ pair will result in a red gauge charge, $e_R$. By an identical argument, fusing the pair of $m_G$ back together will also result in $e_R$.

We now provide more technical details on the derivation of the data corresponding to the twisted quantum double \cite{propitius_topological_1995}. Here, we consider the twisted quantum double $\mathcal D^\alpha(\ZZ_2^3)$, where $\alpha \in H^3(\ZZ_2^3,U(1))$ is a 3-cocycle. Conveniently, we represent the generators of the group $A=\ZZ_2^3$ by order two elements $R,G,B$. Thus, any group element $a \in A$ can be represented as $a=R^{\rho_a} G^{\gamma_a}B^{\beta_a}$ for $\rho_a,\gamma_a,\beta_a \in \{0,1\}$.  
 
 For the twisted quantum double of an Abelian group $A$, an anyon can be labeled by a pair $(a,\sigma)$ where $a \in A$ and $\sigma$ is a choice of projective irreducible representation corresponding to a 2-cocycle $\omega_a$. Such a representation $\Gamma^\sigma_{a}$ satisfies
\begin{align}
    \Gamma^\sigma_{a}(b) \Gamma^\sigma_{a}(c) = \omega_a(b,c)  \Gamma^\sigma_a(bc),
    \label{eq:Gammaprojective}
\end{align}
 and the 2-cocycle $\omega_a$ is related to the input 3-cocycle $\alpha$ via
 \begin{align}
     \omega_a(b,c) =\frac{\alpha(a,b,c)\alpha(b,c,a)}{\alpha(b,a,c)}.
 \end{align}

In particular, our 3-cocycle of interest is given by
\begin{align}
    \alpha(a,b,c) = (-1)^{\rho_a\gamma_b\beta_c}
\end{align}
which implies that
\begin{align}
    \omega_a(b,c) = (-1)^{\rho_a\gamma_b\beta_c + \gamma_a\rho_b\beta_c + \beta_a\rho_b\gamma_c }
\end{align}

We now discuss the possible anyons for each choice of group element $a\in A$

\begin{enumerate}
    \item $a=1$ ($(\rho_a,\gamma_a,\beta_a)= (0,0,0))$). In this case, the 2-cocycle $\omega_a$ is trivial. Therefore, $\Gamma^\sigma_1$ corresponds to (linear) irreducible representations of $\ZZ_2^3$, which correspond exactly to the choice of an element in $A$. We can therefore label such choices as $\sigma\in A$. These correspond to the eight Abelian charges.
    \item $a=R$ ($(\rho_a,\gamma_a,\beta_a)= (1,0,0))$). We find the 2-cocycle $\omega_R(b,c) = (-1)^{\gamma_b\beta_c}$. Eq.~\eqref{eq:Gammaprojective} then implies that $\Gamma_R(G)$ and $\Gamma_R(B)$ anticommute. There are two possible choices of $\Gamma_R$, which we label by $\sigma = \pm$
\begin{align}
    \Gamma^{\pm}_R(R) &= \pm I, & \Gamma^{\pm}_R(G) &= X, & \Gamma^{\pm}_R(B) &= Z,
\end{align}
where $X,Y,Z$ are Pauli matrices and $I$ is the $2\times2$ identity matrix. The two choices correspond to the non-Abelian flux $m_R$ and $f_R$, respectively. Similarly, by permutation of colors, we can define $m_G,f_G,m_B,f_B$ corresponding to the representation $\Gamma^+_G,\Gamma^-_G,\Gamma^+_B,\Gamma^-_B$, respectively where
\begin{align}
\Gamma^{\pm}_G(R) &= Z, & \Gamma^{\pm}_G(G) &= \pm I, & \Gamma^{\pm}_G(B) &= X,\label{eq:GammaG}\\
\Gamma^{\pm}_B(R) &= X, & \Gamma^{\pm}_B(G) &= Z, & \Gamma^{\pm}_B(B) &= \pm I.\label{eq:GammaB}
\end{align}
\item  $a=RG$ ($(\rho_a,\gamma_a,\beta_a)= (1,1,0))$). We find the 2-cocycle $c_{RG}(b,c) = (-1)^{\gamma_b\beta_c + \rho_b\beta_c}$, which means that $\{\Gamma_{RG}(R),\Gamma_{RG}(B)\} = \{\Gamma_{RG}(G),\Gamma_{rg}(B)\}=0$. There are two inequivalent choices given by
\begin{align}
    \Gamma^{\pm}_{RG}(R) &= \pm X, & \Gamma^{\pm}_{RG}(G) &= X, & \Gamma^{\pm}_{RG}(B) &= Z,
\end{align}
which correspond to $m_{RG}$ and $f_{RG}$, respectively. Permuting the colors gives $m_{GB},f_{GB}$ and $m_{RB},f_{RB}$.
\item $a=RGB$ ($(\rho_a,\gamma_a,\beta_a)= (1,1,1))$). We find the 2-cocycle $\omega_{RG}(b,c) = (-1)^{\gamma_b\beta_c + \rho_b \beta_c +  \rho_b \gamma_c}$, which means that all three pairs anticommute. There are two inequivalent choices given by
\begin{align}
    \Gamma^{\pm}_{RGB}(R) &= \pm X, & \Gamma^{\pm}_{RGB}(G) &= \pm Y, & \Gamma^{\pm}_{RGB}(B) &= \pm Z,
\end{align}
which correspond to $s_{RGB}$ and $\bar s_{RGB}$, respectively.
\end{enumerate}

For an Abelian twisted quantum double, the $S$ and $T$ matrices can be computed via the formula~\cite{coste_finite_2000}
\begin{align}
    S_{(a,\sigma),(b,\tau)} &= \frac{1}{|A|}\chi^\sigma_{a}(b)^* \chi^{\tau}_{b}(a)^* & T_{(a,\sigma),(b,\tau)} = \delta_{a,b}\delta_{\sigma,\tau} \frac{\chi^\sigma_{a}(a)}{\chi^\sigma_{a}(1)}
\end{align}
where $\chi^\sigma_a = \text{Tr}[\Gamma^\sigma_a]$ is the character of the corresponding representation. In particular, we find that
\begin{align}
 8S_{(a,\sigma), (b,\tau)} &= 
 \begin{cases}  
    1 &; a=b=1, \\
    2\times (-1)^{\rho_a\rho_\tau+\gamma_a\gamma_\tau+ \beta_a\beta_\tau} &; a=1, b\ne 1,\\
    4 \times \delta_{a,b}(-1)^{\rho_a \gamma_a \beta_a + \sigma \tau} &; a\ne 1, b\ne 1,
 \end{cases}\\
 \text{diag } T_{(a,\sigma)} &=
  \begin{cases}  
  1&; a=1,\\
 (-1)^{\sigma} \times i^{\rho_a \gamma_a \beta_a} & a \ne 1.
  \end{cases}
\end{align}
where we remind that $\sigma \in A$ when $a =1$ and $\sigma = \pm 1$ for $a\ne 1$.
For completion, the full matrices are
\begin{align}
\begin{array}{ccccccccccccccccccccccc}
&1 & e_R & e_G & e_B & e_{RG} & e_{GB} & e_{RB} & e_{RGB} & m_R & f_R & m_G & f_G & m_B & f_B & m_{RG} & f_{RG} & m_{GB} & f_{GB} & m_{RB} & f_{RB} & s_{RGB} & \bar{s}_{RGB} \\
\text{diag } T =& (1 & 1 & 1 & 1 & 1 & 1 & 1 & 1 & 1 & -1 & 1 & -1 & 1 & -1 & 1 & -1 & 1 & -1 & 1 & -1 & i & -i) \\
\end{array}\end{align}
\begin{align}
S=\frac{1}{8}
\begin{pmatrix}
 1 & 1 & 1 & 1 & 1 & 1 & 1 & 1 & 2 & 2 & 2 & 2 & 2 & 2 & 2 & 2 & 2 & 2 & 2 & 2 & 2 & 2 \\
 1 & 1 & 1 & 1 & 1 & 1 & 1 & 1 & -2 & -2 & 2 & 2 & 2 & 2 & -2 & -2 & 2 & 2 & -2 & -2 & -2 & -2 \\
 1 & 1 & 1 & 1 & 1 & 1 & 1 & 1 & 2 & 2 & -2 & -2 & 2 & 2 & -2 & -2 & -2 & -2 & 2 & 2 & -2 & -2 \\
 1 & 1 & 1 & 1 & 1 & 1 & 1 & 1 & 2 & 2 & 2 & 2 & -2 & -2 & 2 & 2 & -2 & -2 & -2 & -2 & -2 & -2 \\
 1 & 1 & 1 & 1 & 1 & 1 & 1 & 1 & -2 & -2 & -2 & -2 & 2 & 2 & 2 & 2 & -2 & -2 & -2 & -2 & 2 & 2 \\
 1 & 1 & 1 & 1 & 1 & 1 & 1 & 1 & 2 & 2 & -2 & -2 & -2 & -2 & -2 & -2 & 2 & 2 & -2 & -2 & 2 & 2 \\
 1 & 1 & 1 & 1 & 1 & 1 & 1 & 1 & -2 & -2 & 2 & 2 & -2 & -2 & -2 & -2 & -2 & -2 & 2 & 2 & 2 & 2 \\
 1 & 1 & 1 & 1 & 1 & 1 & 1 & 1 & -2 & -2 & -2 & -2 & -2 & -2 & 2 & 2 & 2 & 2 & 2 & 2 & -2 & -2 \\
 2 & -2 & 2 & 2 & -2 & 2 & -2 & -2 & 4 & -4 & 0 & 0 & 0 & 0 & 0 & 0 & 0 & 0 & 0 & 0 & 0 & 0 \\
 2 & -2 & 2 & 2 & -2 & 2 & -2 & -2 & -4 & 4 & 0 & 0 & 0 & 0 & 0 & 0 & 0 & 0 & 0 & 0 & 0 & 0 \\
 2 & 2 & -2 & 2 & -2 & -2 & 2 & -2 & 0 & 0 & 4 & -4 & 0 & 0 & 0 & 0 & 0 & 0 & 0 & 0 & 0 & 0 \\
 2 & 2 & -2 & 2 & -2 & -2 & 2 & -2 & 0 & 0 & -4 & 4 & 0 & 0 & 0 & 0 & 0 & 0 & 0 & 0 & 0 & 0 \\
 2 & 2 & 2 & -2 & 2 & -2 & -2 & -2 & 0 & 0 & 0 & 0 & 4 & -4 & 0 & 0 & 0 & 0 & 0 & 0 & 0 & 0 \\
 2 & 2 & 2 & -2 & 2 & -2 & -2 & -2 & 0 & 0 & 0 & 0 & -4 & 4 & 0 & 0 & 0 & 0 & 0 & 0 & 0 & 0 \\
 2 & -2 & -2 & 2 & 2 & -2 & -2 & 2 & 0 & 0 & 0 & 0 & 0 & 0 & 4 & -4 & 0 & 0 & 0 & 0 & 0 & 0 \\
 2 & -2 & -2 & 2 & 2 & -2 & -2 & 2 & 0 & 0 & 0 & 0 & 0 & 0 & -4 & 4 & 0 & 0 & 0 & 0 & 0 & 0 \\
 2 & 2 & -2 & -2 & -2 & 2 & -2 & 2 & 0 & 0 & 0 & 0 & 0 & 0 & 0 & 0 & 4 & -4 & 0 & 0 & 0 & 0 \\
 2 & 2 & -2 & -2 & -2 & 2 & -2 & 2 & 0 & 0 & 0 & 0 & 0 & 0 & 0 & 0 & -4 & 4 & 0 & 0 & 0 & 0 \\
 2 & -2 & 2 & -2 & -2 & -2 & 2 & 2 & 0 & 0 & 0 & 0 & 0 & 0 & 0 & 0 & 0 & 0 & 4 & -4 & 0 & 0 \\
 2 & -2 & 2 & -2 & -2 & -2 & 2 & 2 & 0 & 0 & 0 & 0 & 0 & 0 & 0 & 0 & 0 & 0 & -4 & 4 & 0 & 0 \\
 2 & -2 & -2 & -2 & 2 & 2 & 2 & -2 & 0 & 0 & 0 & 0 & 0 & 0 & 0 & 0 & 0 & 0 & 0 & 0 & -4 & 4 \\
 2 & -2 & -2 & -2 & 2 & 2 & 2 & -2 & 0 & 0 & 0 & 0 & 0 & 0 & 0 & 0 & 0 & 0 & 0 & 0 & 4 & -4 \\
\end{pmatrix} 
\end{align}

Using the S-matrix above, the fusion rules in Table \ref{tab:D4fusion} can be derived using the Verlinde formula~\cite{verlinde_fusion_1988}
\begin{align}
   N^{K}_{IJ} = \sum_{L}\frac{S_{IL}S_{JL}S_{KL}^*}{S_{1L}}
\end{align}
where $N^{K}_{IJ}$ are fusion multiplicities for the process $I \times J = \sum_K N^{K}_{IJ} K$, and $I,J,K,L$ are anyons.

Next, let us demonstrate that by creating a pair of $m_G$ and $m_B$ anyons, braiding them and fusing each pair back together, we are left with two $e_R$'s as fusion outcomes. To be concrete, we will consider creating a pair of $m_G$ at positions $1$ and $2$ and a pair of $m_B$ at positions $3$ and $4$. We then perform a full braid of $m_G$ at position 2 and $m_B$ at position 3, before fusing each pair back together. Note that because each flux has quantum dimension $2$, the internal state forms a qubit on which the representations $\Gamma$ act on.

To create a pair of $m_G$ fluxes, we must ensure that the state transforms under the trivial representation of $\Gamma^+_G \otimes \Gamma^+_G$. From Eq.~\ref{eq:GammaG}, we see that the symmetry action of red, green and blue on the two sites given by $Z_1Z_2, I_1I_2, X_1X_2$, respectively. Therefore, the ``singlet" state under this symmetry action is stabilized $Z_1Z_2$ and $X_1X_2$ (i.e. the Bell state $\frac{1}{\sqrt{2}}(\ket{\uparrow_1 \uparrow_2}+ \ket{\downarrow_1 \downarrow_2})$. Similarly, a pair of $m_B$ anyons created from the vacuum is stabilized by $X_3X_4$ and $Z_3Z_4$.

In general, the full braid of an anyon $(a,\sigma)$ around $(b,\tau)$ is given by the unitary operation $\Gamma^\sigma(b)_a \times \Gamma^\tau_b(a)$. In this case, we have $(a,\sigma) = (G,+)$ and $(b,\tau) = (B,+)$. Therefore, the braiding operation on sites $2$ and $3$ is implemented via $\Gamma^+_G(B) \times \Gamma^+_B(G) = X_2Z_3$. Thus, we see qualitatively that we are toggling the fusion channel space of the $m_G$ pair by $X$, as described in the main text. To see this explicitly, conjugating the original stabilizers, we see that the state after the braiding is stabilized by $-Z_1Z_2$, $X_1X_2$, $Z_3Z_4$, $-X_3X_4$. In particular, the state satisfies $\Gamma^+_G(R)_1 \Gamma^+_G(R)_2 = Z_1Z_2 = -1$, which means it is charged under the red symmetry action. This signifies that the fusion outcome of the two $m_G$ anyons will result in $e_R$. Similarly, $\Gamma^+_B(R)_3 \Gamma^+_B(R)_4 = X_3X_4 = -1$, implying that the two $m_B$ anyons also fuse into $e_R$.

We can also demonstrate that the Borromean ring braiding of $m_R$, $m_G$ and $m_B$ results in a phase $-1$. The Borromean ring braiding can be deformed topologically to the following process~\cite{wang_topological_2015}
\begin{enumerate}
    \item Create $m_R$, $m_G$, $m_B$ pairs at sites $1$ and $2$, $3$ and $4$, $5$ and $6$, respectively.
    \item Braid sites $2$ and $4$ followed by $2$ and $6$
    \item Braid (in the reverse direction) sites $2$ and $4$ followed by $2$ and $6$
    \item Fuse each flux pair.
\end{enumerate}
Braiding sites $2$ and $4$ is realized by the operation $\Gamma^+_R(G)_2 \Gamma^+_G(R)_4 = X_2 Z_4$ while braiding sites $2$ and $6$ is realized by $\Gamma^+_R(B)_2 \Gamma^+_B(R)_6 = Z_2 X_6$. Therefore, the total braiding is realized by the operation
\begin{align}
 (Z_2 X_6)^{-1} (X_2 Z_4)^{-1} (Z_2 X_6) (X_2 Z_4) = -1
\end{align}

\subsection{Non-square degeneracy}
\label{methods_nonsquare_GSD}
We provide a simple argument that all Abelian anyon theories that admit a gapped boundary (which excludes for example, chiral phases) have a perfect square ground state degeneracy on a torus. If there exists a gapped boundary, then there exists a Lagrangian subgroup $K$ of bosons which have trivial mutual braiding statistics and braids non-trivially with all other anyons outside this group. For Abelian anyon theories, $|K|^2$ is equal to the the total number of anyons(see e.g.~\cite{Kaidi22}), which is also equal to the ground state degeneracy on a torus.

\subsection{Qubit Reuse and Circuit Optimization Techniques}
\label{appendix_circuit_optimisation}
Here we describe the steps we need to realize the $D_4$ topological order on  the H2 trapped ion device. The construction requires a total of $27+3=30$ physical qubits to create 27-qubit states.
During the procedure 9 ancillas need to be measured for a layer of feed-forward. It is crucial to reuse some of those ancilla qubits to fit the protocol into the maximal qubit capacity (32) of the device.
We begin by rewriting Eq.~(\ref{eq_gs_prep}) as
\begin{equation}	
	\ket{\psi_0} = \prod_v H_v \bra{+}_P  \prod_{\braket{v,p}} CZ_{v,p}
	\!\!\!\prod_{\langle p,{\tilde p},\tilde{\tilde p}\rangle}\!\!\! e^{\pm \frac{i \pi}{8} Z_{p}Z_{{\tilde p}}Z_{{\tilde{\tilde p}}}}  \ket{+}_{P}\ket{+}_V,
	\label{eq_gs_prep_2}
\end{equation}
where $|+\rangle_P = |\red{+}\rangle_{\red{P}} |\green{+}\rangle_{\green{P}} |\blue{+}\rangle_{\blue{P}}$ 
and $|+\rangle_V = |\red{+}\rangle_{\red{V}} |\green{+}\rangle_{\green{V}} |\blue{+}\rangle_{\blue{V}}$ denote plaquette and vertex qubits over the red, green and blue sublattices. Moreover, the product over $CZ$s also spans over three sublattices which can be decoupled, i.e,
$$\prod_{\braket{v,p}} CZ_{v,p}=\prod_{\braket{\red{v,p}}} CZ_{\red{v,p}} \prod_{\braket{\green{v,p}}} CZ_{\green{v,p}} \prod_{\braket{\blue{v,p}}} CZ_{\blue{v,p}}.$$ 
We can rearrange Eq.~(\ref{eq_gs_prep_2}) as 
\begin{equation}	
	\ket{\psi_0}\! = \prod_v\! H_v \!
	\bra{\red{+}}_{\red{P}}\!\!  \prod_{\braket{\red{v,p}}}\! CZ_{\red{v,p}} \ket{\red{+}}_{\red{V}}
	\underbrace{
	\bra{\green{+}}_{\green{P}} \bra{\blue{+}}_{\blue{P}}\!\!\!\!
	\prod_{\langle p,{\tilde p},{\tilde{\tilde p}}\rangle}\!\!\!\! e^{\pm \frac{i \pi}{8} Z_{p}Z_{{\tilde p}}Z_{{\tilde {\tilde p}}}}  |\red{+}\rangle_{\red{P}}\!\!
	\left(\prod_{\braket{\green{v,p}}}\! CZ_{\green{v,p}}  |\green{+}\rangle_{\green{P}} \ket{\green{+}}_{\green{V}} \right)\!\!\!
	\left(\prod_{\braket{\blue{v,p}}}\! CZ_{\blue{v,p}}|\blue{+}\rangle_{\blue{P}}\ket{\blue{+}}_{\blue{V}}
	\!\!\right)\!\!.
	}_{\text{step 1}}
	\label{eq_gs_prep_3}
\end{equation}
The construction is divided into two steps. In step 1, we act on the vertex qubits of the green and blue sublattices only, and all the plaquette qubits of all three sublattices.
In total, we use $(\green{9}+\blue{9})=18$ vertex and $(\red{3}+\green{3}+\blue{3})=9$ plaquette qubits during step 1. At the end of step 1, we measure $(\green{3}+\blue{3})=6$ plaquette qubits of the green and blue sublattices. During step 2, we reuse the 6 measured qubits as vertex qubits of the red sublattice and act with $CZ$ gates before also measuring the plaquette qubits of the red sublattice.

During step 1, we choose to first implement the $CZ$ gates between the vertices and plaquette qubit. For a given color, we have $3+3=6$ CZ gates between 3 vertex and 2 plaquette qubits (e.g., we have 3 $CZ$s between $v_1$, $v_2$, $v_3$, and $p_1$, and 3 $CZ$s between the same vertices and $p_2$). Since $CZ$ gates for the blue and green sublattices act on qubits in the $|+\rangle$ state, the action of the later 3 $CZ$ gates can be implemented by a single two-qubit gate. This can be seen by the state-vector identity for e.g., the blue sublattice,
\begin{equation}
	\begin{aligned}
		\cbox{13}{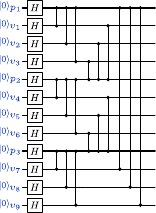}  &= \cbox{33}{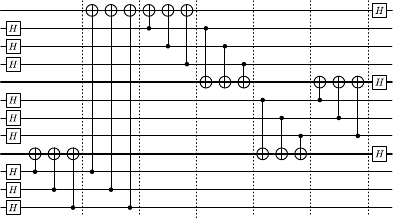}\\
		&= \cbox{33}{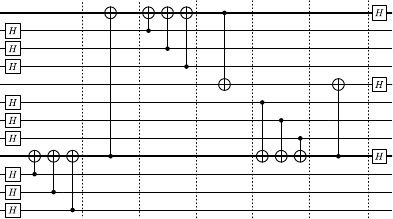}
	\end{aligned}
\end{equation}
Similar identities also holds for the green sublattice. This leads to a reduction from $3\times6\times2=36$ to $3\times4\times2=24$ two-qubit gates for the products of $CZ$ on the blue and green sublattices.

\begin{figure*}[!ht]
\centering
\includegraphics[width=1.0\textwidth]{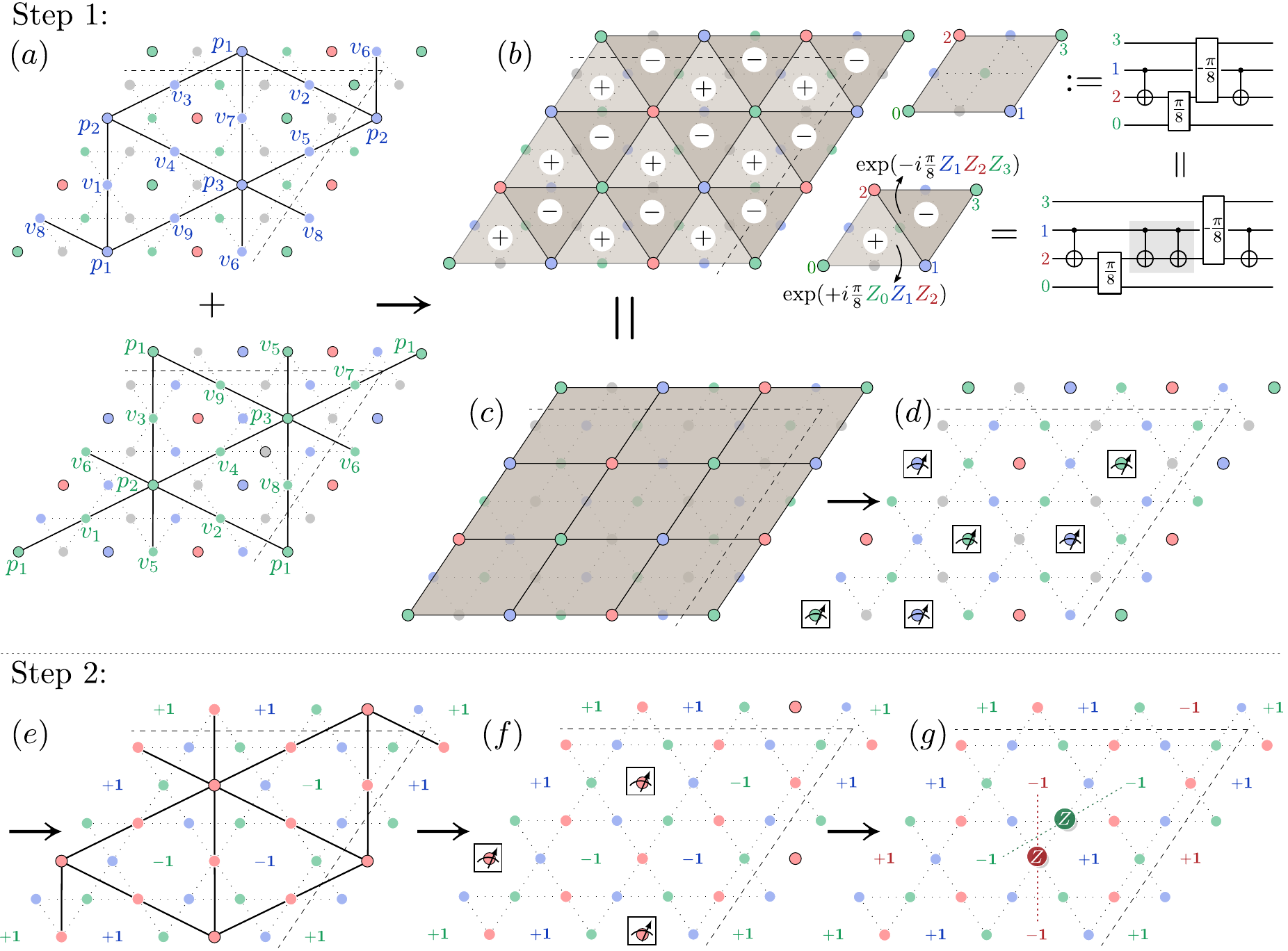}
\caption{\textbf{More detail on the steps involved in preparing the ground state}. In step 1a, we apply $CZ$ gates on the blue and green sublattices, using the circuit optimisation techniques from section~\ref{appendix_circuit_optimisation}. (b) shows the implementation of two adjacent $\exp({\pm \frac{i \pi}{8} Z_{p}Z_{\tilde{p}}Z_{\tilde{\tilde{p}}}})$ gates within each square using 4 two-qubit gates. (c) as shown in (b), two adjecent $\exp({\pm \frac{i \pi}{8} Z_{p}Z_{\tilde{p}}Z_{\tilde{\tilde{p}}}})$ gates within each square can be combined into one parallelogram of four two-qubit gates. (d) we measure the plaquette qubits of green and blue sublattices. In step 2e,  we apply $CZ$ gates between vertex and plaquette qubits of the red sublattice. (f) we measure plaquette qubits of the red sublattice. (g) shows a feed-forward action determined by the outcome of red, green and blue plaquette qubits.}
%\textbf{Caption.} This is how we actually do it... \hd{Needs a proper caption - after updating the style of all figures}
\label{fig_app_preparation}
\end{figure*}

The $\exp({\pm \frac{i \pi}{8} Z_{p}Z_{{\tilde p}}Z_{{\tilde{\tilde p}}}})$ gate that acts on 3 plaquette qubits of distinct colors is decomposed as
\begin{equation}
		e^{\pm \frac{i \pi}{8} Z_{p}Z_{{\tilde p}}Z_{{\tilde{\tilde p}}}}= \cbox{8}{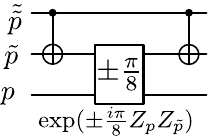}
		\label{eq_exp_zzz}
\end{equation}
This decomposition uses the parametrised entangling gate ZZPhase$(\theta)=e^{-i\theta/2Z\otimes Z}$ which is the native entangling gate on the device. Without any optimization, we need $18\times 3=54$ gates to realize the action of $\exp({\pm \frac{i \pi}{8} Z_{p}Z_{\tilde p}Z_{\tilde{\tilde{p}}}})$ gates. Since the order of qubits in (\ref{eq_exp_zzz}) does not matter, we notice that the stacked action of $\exp({+ \frac{i \pi}{8} Z_{0}Z_{1}Z_{2}})$ and $\exp({- \frac{i \pi}{8} Z_{1}Z_{2}Z_{3}})$ requires not 6 but 4 two-qubit gates (Extended Data Fig. \ref{fig_app_preparation}). By using this `squared' implementation of the triangular $\exp({\pm \frac{i \pi}{8} Z_{p}Z_{\tilde{p}}Z_{\tilde{\tilde{p}}}})$ gates we reduce the number of two-qubits gates for this step from 54 to $9\times4=36$.

By using these circuit optimizations, we reduce the gate count to $\red{18}+\green{12}+\blue{12}=42$ two qubit gates for the `$CZ$-chains' and $36$ two-qubit gates for the $\exp({\pm \frac{i \pi}{8} Z_{p}Z_{\tilde{p}}Z_{\tilde{\tilde{p}}}})$ gates which give us a total of $42+36=78$ two-qubit gates.

We used TKET for compiling circuits into native gates \cite{sivarajah_tket_2021}. The resulting circuit (for state preparation only) consisted of $165$ one-qubit gates, 60 ZZMax $= e^{-i\pi/4Z\otimes Z}$ and 18 ZZPhase gates, while the depth of the native circuit is 56.

%\putbib

\section*{Data availability}
The numerical data that support the findings of this study are available from Zenodo repository 10.5281/zenodo.7848424~\cite{iqbal_supporting_2023}.

\section*{Code availability}
The code used for numerical simulations is available from from Zenodo repository 10.5281/zenodo.7848424~\cite{iqbal_supporting_2023}.

\section*{Acknowledgements}
We thank the broader team at Quantinuum for discussions,  feedback and encouragement, especially David Hayes, Konstantinos Meichanetzidis, Luuk Coopmans, Yuta Kikuchi, Patty Lee and Ilyas Khan. R.V. thanks Nick Jones and Rahul Sahay for comments on the manuscript.
N.T. is supported by the Walter Burke Institute for Theoretical Physics at Caltech. R.V. is supported by the Harvard Quantum Initiative Postdoctoral Fellowship in Science and Engineering. A.V. and R.V. are supported by the Simons Collaboration on Ultra-Quantum Matter, which is a grant from the Simons Foundation (618615, A.V.). The experimental data in this work was produced by the Quantinuum H2 trapped ion quantum computer, powered by Honeywell.

\section*{Author contributions}
M.I. wrote the code generating the circuits for all experiments. The experiment was built and carried out by S.L.C., J.M.D., C.F., J.P.G., J.J., M.M., S.A.M., J.M.P., A.R., M.R.,P.S. and R.P.S. 
The data analysis and interpretation was done by M.I., N.T, M.F., A.V, R.V. and H.D.. N.T., A.V. and R.V. contributed to the ideation, theory and experiment design, including the definition of the operators for anyon creation, movement and annihilation. H.D. drafted the initial manuscript, which was refined by contributions from all authors, especially M.I., N.T., R.V. and A.V.

\section*{Competing interests} H.D. is a shareholder of Quantinuum. All other authors declare no competing interests.

\section*{Additional information}
Correspondence and requests for materials should be addressed to H.D. at henrik.dreyer (at) quantinuum.com.

\clearpage
\section{Extended Data Figures and Tables}

\clearpage

\clearpage

\clearpage

%\begin{figure*}[!ht]
%	\centering
%\includegraphics[width=0.55\textwidth]{appendix_pdfs/10body.pdf}
	%\caption{\label{edf_10body}\textbf{Data analysis with 10-body star terms.}
  %The model~(\ref{eq_hamiltonian}) has the same ground states when one deletes all but two $CZ$ operators in each $A_s$ on opposite sides of a hexagon.
  %This figure shows the data with respect to the modified star operators and is computed by discarding heralded shots without any error mitigation (same as setting (b) in Extended Data Fig.~\ref{fig_spam_comparison}).}
	%\label{fig_10body}
%\end{figure*}

\begin{figure*}[!ht]
	\centering
	\includegraphics[width=0.95\textwidth]{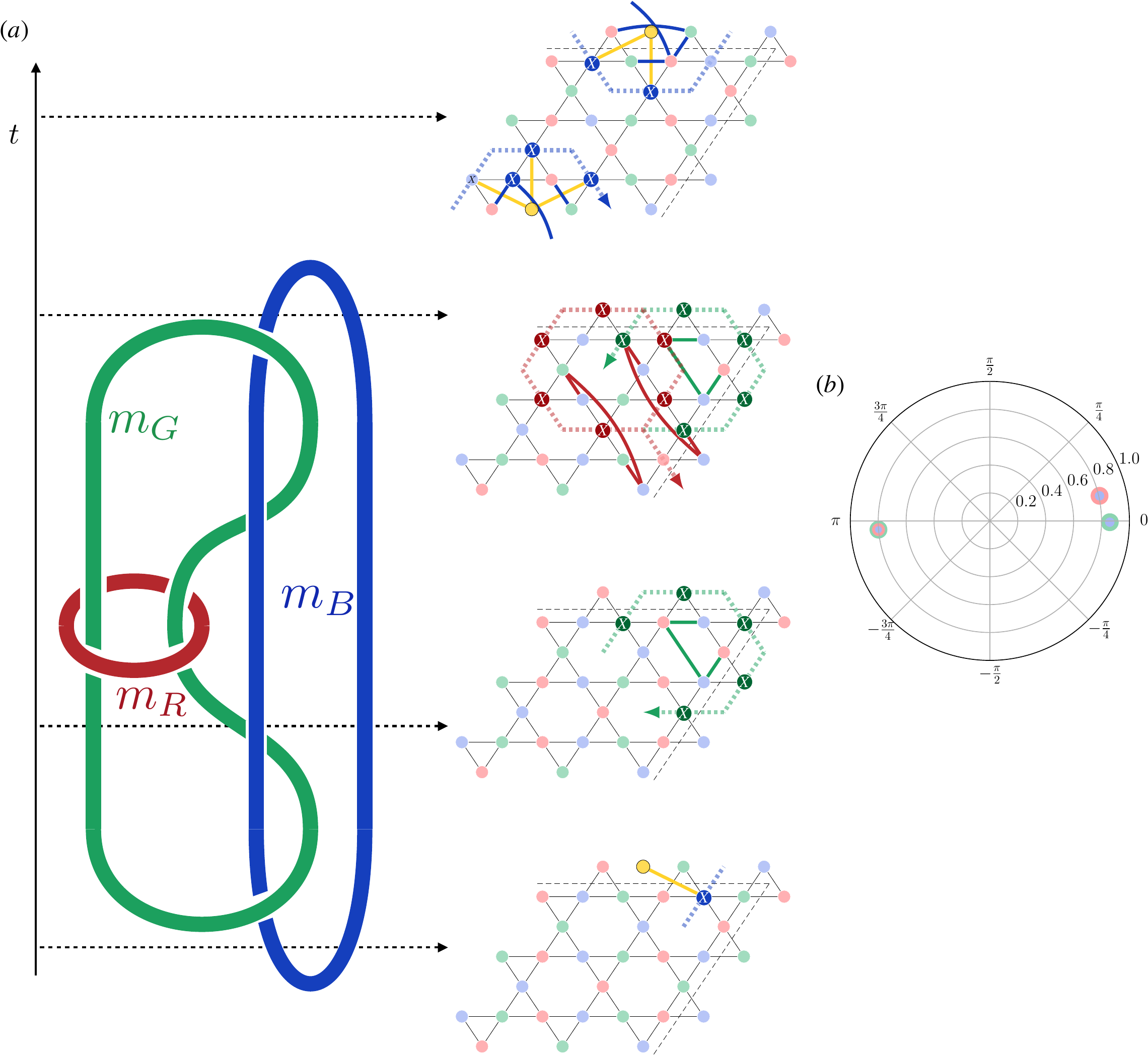}
	\caption{\textbf{Details on the implementation of the Borromean Rings.} (a) Going up in time the thumbnails on the right show all gates applied during the braiding sequence of the Borromean rings (cf. section~\ref{sec_braiding}). The solid colored lines are $CZ$ gates and the dashed black lines denote the periodic boundary conditions. The yellow ancilla qubit is controlling \emph{all} blue gates including both $X$ and $CZ$. After the ground state~(\ref{eq_gs_prep}) is prepared, the ancilla is initialised and measured in the $X$- and  $Y$-basis (in different shots) to extract both the real and the imaginary part of the phase. Since only the trajectory of the blue non-Abelion pair is controlled, this measures the phase of a spacetime diagram that is identical to the one shown on the left \emph{plus} a disconnected diagram where the blue loop is missing. However, since the red and green loops in that disconnected part of the diagram are contractible, this is topologically equivalent to the Borromean Rings. (b) Measured value for the phase of the braid containing all three colors, as well as the braids where either red or green are removed. Writing  $\braket{\psi_0| \text{braid} | \psi_0} = r e^{i\phi}$, the measured values for the different braids are: \red{red}-\green{green}-\blue{blue}: $r=0.80(2)$, $\phi = 1.02(2)\pi$, \red{red}-\blue{blue}: $r=0.81(2)$, $\phi = 0.07(2)\pi$, \green{green}-\blue{blue}: $r=0.86(2)$, $\phi = 2.00(2)\pi$. Uncertainty is one standard error on the mean.
 }
	\label{EDF_borromean}
\end{figure*}
\clearpage

\begin{figure*}[!ht]
	\centering
	\includegraphics[width=0.95\textwidth]{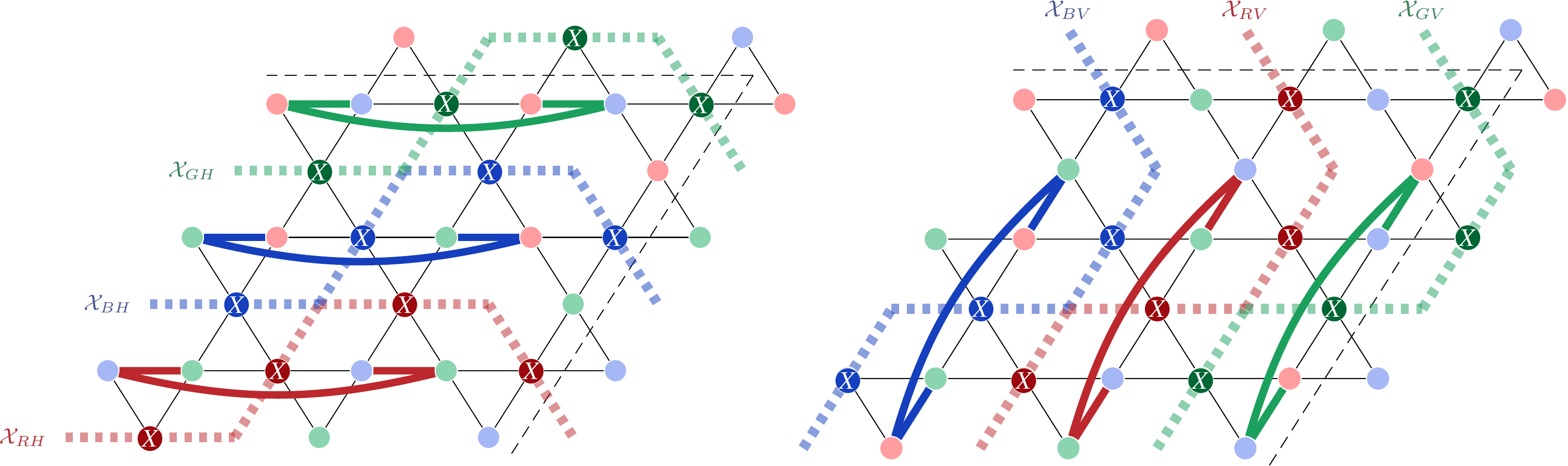}
	\caption{\textbf{Non-contractible loops applied to switch logical sectors.} Operators applied to obtain all 22 ground states of the model in section ~\ref{sec_noncontractible_loops}, specifically Fig.~\ref{fig_23gs}. Solid lines denote $CZ$s, and dashed black lines denote the periodic boundary conditions. In the experiment, horizontal strings are applied before vertical strings. If one is interested in the highest possible fidelity state in a given sector, one may alternatively use the protocol~(\ref{eq_gs_prep}) with a computational basis state in the desired sector.
 }
	\label{EDF_loop_geometry}
\end{figure*}
\clearpage

\begin{table*}[h!]
    \centering
    \begin{tabular}{|c|c|c|c|}
    \hline
       &  $m_{R}$ & $m_{RG}$ & $s_{RGB}$  \\
       \hline
    $e$       & $m_R \times e_R = f_R$ & $m_{RG} \times e_R = m_{RG} \times e_G = f_{RG}$ &$s_{RGB} \times e_R =  s_{RGB} \times e_{RGB} = \bar s_{RGB}$\\
     & $m_R \times e_G = m_R \times e_B =m_R$ &$m_{RG} \times e_{RG} = m_{RG} \times e_B =m_{RG}$& $s_{RGB} \times e_{RG} =s_{RGB}$\\
    \hline
    $m_{R}$  &$m_R \times m_R = (1+e_G)(1+e_B)$& $m_{RG} \times m_{G}= m_R + f_R$& $s_{RGB} \times m_R = m_{GB} + f_{GB}$\\
     & $m_{R} \times m_{G} = m_{RG} + f_{RG}$ &$m_{RG} \times m_B = s_{RGB}+ \bar s_{RGB}$ & \\
    \hline
    $m_{RG}$&&$m_{RG} \times m_{RG} = (1+e_{RG})(1+e_B)$&$ s_{RGB} \times m_{RG} = m_{B} + f_{B}$\\
    \hline
    $s_{RGB}$ & &&$s_{RGB} \times s_{RGB} = 1 + e_{RG}+ e_{GB} + e_{RB}$\\
    \hline
    \end{tabular}
    \caption{Summary of fusion rules up to permutation of colors}
    \label{tab:D4fusion}
\end{table*}

\begin{table}[h!]
    \centering
        \begin{tabular}{|c|c|c||c||c|c|}
        \hline
 \multicolumn{3}{|c||}{$\mathcal D(D_4)$} & \multirow{2}{*}{$\mathcal D^\alpha(\ZZ_2^3)$} & \multirow{2}{*}{dim.} & \multirow{2}{*}{$T$}\\
 \cline{1-3}
Conj. class & Centralizer & irrep &&  &\\
    \hline
          $1$ &$D_4$ & $\bs 1$  & 1      & 1 & 1\\
      $1$ &$D_4$ &  $\bs s_1$  & $e_{RG}$      & 1  & 1\\
        $1$ &$D_4$ &  $\bs s_2$ & $e_{R}$      & 1 & 1\\
  $1$ &$D_4$ & $\bs s_3$  & $e_{G}$       & 1 & 1\\
        $1$ &$D_4$ & $\bs 2$  & $m_B$       & 2 & 1 \\
\hline
          ${r^2}$ &$D_4$ &$\bs 1$  & $e_{RGB}$   & 1 & 1\\
${r^2}$ &$D_4$ &$\bs s_1$  & $e_{B}$   & 1 & 1\\
     ${r^2}$ &$D_4$ &  $\bs s_2$ &$e_{GB}$    & 1 & 1\\
  ${r^2}$ &$D_4$ &$\bs s_3$ &$e_{RB}$  & 1 & 1\\
      ${r^2}$ &$D_4$ &$\bs 2$  &$f_B$ & 2 & -1\\
\hline
        ${r}$ &$\ZZ_4$ & $\bs 1$ & $m_{RG}$  & 2 & 1\\
             ${r}$ &$\ZZ_4$ & $\bs \omega$ & $ s_{RGB}$  & 2 &$i$\\
   ${r}$ &$\ZZ_4$ & $\bs \omega^2$ & $f_{RG}$  &2 & -1\\
   ${r}$ &$\ZZ_4$ & $\overline{\bs \omega}$ & $\bar s_{RGB}$ & 2 &-$i$\\
\hline
               ${s}$ &$\ZZ_2^2$ & $\bs 1$ &$m_{GB}$ & 2 & 1\\
              ${s}$ &$\ZZ_2^2$ &$(-1,1)$&$m_{G}$ & 2 &1\\
     ${s}$ &$\ZZ_2^2$ & $(1,-1)$ &$f_{G}$& 2 &-1\\
       ${s}$ &$\ZZ_2^2$ & $(-1,-1)$ &$f_{GB}$& 2 &-1 \\
\hline
         ${rs}$ &$\ZZ_2^2$ &$\bs 1$ &$m_{RB}$ & 2 & 1\\
     ${rs}$ &$\ZZ_2^2$ & $(-1,1)$ &$m_{R}$& 2 &1 \\
    ${rs}$ &$\ZZ_2^2$ &$(1,-1)$&$f_{R}$& 2 &-1 \\
 ${rs}$ &$\ZZ_2^2$ & $(-1,-1)$& $f_{RB}$& 2 & -1\\
    \hline
    \end{tabular}
    \caption{Equivalence between $D_4$ anyons and twisted $\ZZ_2^3$ anyons. Anyons of the $D_4$ quantum double model can be classified by a choice of conjugacy class and an irreducible representation of the corresponding centralizer. Here, the group $D_4$ is generated by group elements $r$ and $s$, which satisfy $r^4 = s^2 = (rs)^2 =1$.}
        \label{tab:D4anyons}
    \end{table}

\clearpage

\input{table-1.tex}

\input{table-2.tex}

\end{document}

%% file: tikz/colorscheme.tex
\definecolor{lightred}{RGB}{255,205,207}
\definecolor{lightgreen}{RGB}{205,255,230}
\definecolor{lightblue}{RGB}{205,216,255}

\definecolor{midred}{RGB}{255,157,160}
\definecolor{midgreen}{RGB}{139,212,176}
\definecolor{midblue}{RGB}{166,183,245}

\definecolor{darkred}{RGB}{155,8,13}
\definecolor{darkgreen}{RGB}{12,110,61}
\definecolor{darkblue}{RGB}{21,63,189}
\newcommand{\red}[1]{\textcolor{darkred}{#1}}
\newcommand{\green}[1]{\textcolor{darkgreen}{#1}}
\newcommand{\blue}[1]{\textcolor{darkblue}{#1}}

%% file: table-1.tex
    \begin {table*}
    \begin{center}
    	\bgroup
    	\def\arraystretch{1.2}% 
    	\scalebox{0.85}{
    		\begin{tabular}{c @{\ }@{\ }|@{\ }@{\ } c } 
    			%		\hline
    			Ground State &
    			
    			\begin{tabular}{@{}c@{}}
                $\qquad\qquad\qquad\qquad\qquad\qquad\qquad\qquad\qquad\quad[e^\text{\ding{65}}_1,\ e^{\triangleleft}_1,\ e^{\triangleright}_1],\ 
    				[e^\text{\ding{65}}_2,\ e^{\triangleleft}_2,\ e^{\triangleright}_2],\ \dots 
    				[e^\text{\ding{65}}_9,\ e^{\triangleleft}_9,\ e^{\triangleright}_9]
    				\qquad\qquad\qquad\qquad\qquad\qquad\qquad\qquad\qquad\quad
    				$ \vspace{0.2em}\\
    				\arrayrulecolor{gray}\hline\arrayrulecolor{black}\vspace{0.2em}
    				$
    				{\mathcal{Z}_{RH}},\ {\mathcal{Z}_{GH}},\ 
    				{\mathcal{Z}_{BH}},\ 
    				{\mathcal{Z}_{RV}},\ {\mathcal{Z}_{GV}},\ {\mathcal{Z}_{BV}}
    				$ \\ 
    				$
    				{\mathcal{X}_{RH}},\ {\mathcal{X}_{GH}},\ 
    				{\mathcal{X}_{BH}},\ 
    				{\mathcal{X}_{RV}},\ {\mathcal{X}_{GV}},\ {\mathcal{X}_{BV}}
    				$ 
    				\vspace{0.2em}
    			\end{tabular}
    			
    			\\ \hline\hline					
        
\begin{tabular} {@{}c@{}} $\red{1,}\green{1,}\blue{1,}\red{1,}\green{1,}\blue{1}$ \\ $\overline{e}$ = -0.946(4) \\ $\overline{l}$ = 0.974(1) \\ $\overline{\mathcal{X}}$ = -0.01(1) 
        \vspace{0.2em}
        \end{tabular} & 
        \begin{tabular}{@{}c@{}} 
        $[0.94(1), 0.95(1), 0.97(1)], [0.91(1), 0.979(9), 0.95(1)], [0.91(1), 0.97(1), 0.97(1)], [0.88(2), 0.94(1), 0.94(1)], [0.92(1), 0.95(1), 0.980(8)],$ \\ 
        $[0.91(2), 0.96(1), 0.95(1)], [0.92(1), 0.97(1), 0.988(6)], [0.94(1), 0.97(1), 0.97(1)], [0.90(1), 0.96(1), 0.96(1)]$ \\
        \arrayrulecolor{gray}\hline\arrayrulecolor{black}\vspace{0.2em}
        $0.981(3), 0.975(4), 0.967(5), 0.978(4), 0.973(4), 0.970(5)$ \\ 
        $0.050(2), -0.010(2), -0.023(1), -0.016(2), -0.007(2), -0.029(1)$  
        \vspace{0.2em}
        \end{tabular} 
        \\ \hline
        
\begin{tabular} {@{}c@{}} $\red{\text{-}1,}\green{1,}\blue{1,}\red{1,}\green{1,}\blue{1}$ \\ $\overline{e}$ = -0.937(6) \\ $\overline{l}$ = 0.965(3) \\ $\overline{\mathcal{X}}$ = -0.05(2) 
        \vspace{0.2em}
        \end{tabular} & 
        \begin{tabular}{@{}c@{}} 
        $[0.95(2), 0.97(1), 0.95(2)], [0.90(3), 0.96(2), 0.98(1)], [0.93(2), 0.98(1), 0.99(1)], [0.83(3), 0.97(1), 0.94(2)], [0.89(3), 0.95(2), 0.98(1)],$ \\ 
        $[0.87(3), 0.91(2), 0.90(3)], [0.90(3), 0.94(2), 0.96(1)], [0.94(2), 0.93(2), 0.99(1)], [0.85(3), 0.97(1), 0.97(1)]$ \\
        \arrayrulecolor{gray}\hline\arrayrulecolor{black}\vspace{0.2em}
        $-0.979(6), 0.964(8), 0.960(8), 0.973(7), 0.962(8), 0.954(8)$ \\ 
        $-0.117(4), 0.078(2), 0.025(2), -0.213(4), -0.109(4), 0.065(4)$  
        \vspace{0.2em}
        \end{tabular} 
        \\ \hline
        
\begin{tabular} {@{}c@{}} $\red{1,}\green{\text{-}1,}\blue{1,}\red{1,}\green{1,}\blue{1}$ \\ $\overline{e}$ = -0.949(5) \\ $\overline{l}$ = 0.976(2) \\ $\overline{\mathcal{X}}$ = 0.03(2) 
        \vspace{0.2em}
        \end{tabular} & 
        \begin{tabular}{@{}c@{}} 
        $[0.95(2), 0.98(1), 0.98(1)], [0.89(3), 0.98(1), 0.94(2)], [0.92(2), 0.99(1), 0.95(2)], [0.93(2), 0.99(1), 0.97(1)], [0.86(3), 0.95(2), 0.98(1)],$ \\ 
        $[0.92(2), 0.96(2), 0.95(2)], [0.92(2), 0.98(1), 0.97(1)], [0.90(3), 0.95(2), 0.99(1)], [0.93(2), 0.97(1), 0.97(1)]$ \\
        \arrayrulecolor{gray}\hline\arrayrulecolor{black}\vspace{0.2em}
        $0.982(5), -0.979(5), 0.972(6), 0.977(6), 0.981(5), 0.964(8)$ \\ 
        $0.067(2), 0.007(5), 0.023(2), -0.056(4), 0.082(5), 0.072(4)$  
        \vspace{0.2em}
        \end{tabular} 
        \\ \hline
        
\begin{tabular} {@{}c@{}} $\red{1,}\green{1,}\blue{\text{-}1,}\red{1,}\green{1,}\blue{1}$ \\ $\overline{e}$ = -0.925(7) \\ $\overline{l}$ = 0.965(3) \\ $\overline{\mathcal{X}}$ = 0.02(2) 
        \vspace{0.2em}
        \end{tabular} & 
        \begin{tabular}{@{}c@{}} 
        $[0.88(3), 0.96(2), 0.94(2)], [0.90(3), 0.97(1), 0.98(1)], [0.88(3), 0.92(2), 0.97(1)], [0.92(2), 0.98(1), 0.99(1)], [0.91(3), 0.92(2), 0.98(1)],$ \\ 
        $[0.81(4), 0.87(3), 0.93(2)], [0.91(3), 0.94(2), 0.95(2)], [0.82(4), 0.91(2), 0.96(2)], [0.89(3), 0.97(1), 0.97(1)]$ \\
        \arrayrulecolor{gray}\hline\arrayrulecolor{black}\vspace{0.2em}
        $0.981(6), 0.960(8), -0.965(7), 0.981(5), 0.94(1), 0.961(7)$ \\ 
        $-0.041(2), 0.009(2), 0.060(4), 0.016(4), 0.016(4), 0.064(4)$  
        \vspace{0.2em}
        \end{tabular} 
        \\ \hline
        
\begin{tabular} {@{}c@{}} $\red{1,}\green{1,}\blue{1,}\red{\text{-}1,}\green{1,}\blue{1}$ \\ $\overline{e}$ = -0.952(5) \\ $\overline{l}$ = 0.974(2) \\ $\overline{\mathcal{X}}$ = -0.01(2) 
        \vspace{0.2em}
        \end{tabular} & 
        \begin{tabular}{@{}c@{}} 
        $[0.93(2), 0.97(1), 0.94(2)], [0.91(3), 0.95(2), 0.97(1)], [0.88(3), 0.99(1), 0.96(2)], [0.92(2), 0.96(2), 0.96(2)], [0.92(2), 0.93(2), 0.99(1)],$ \\ 
        $[0.98(1), 0.97(1), 0.98(1)], [0.94(2), 0.94(2), 0.97(1)], [0.97(1), 0.93(2), 1.0], [0.90(3), 0.98(1), 0.99(1)]$ \\
        \arrayrulecolor{gray}\hline\arrayrulecolor{black}\vspace{0.2em}
        $0.976(6), 0.970(7), 0.975(6), -0.974(6), 0.970(7), 0.979(6)$ \\ 
        $-0.067(5), -0.135(4), -0.066(4), 0.126(5), 0.050(2), 0.049(3)$  
        \vspace{0.2em}
        \end{tabular} 
        \\ \hline
        
\begin{tabular} {@{}c@{}} $\red{1,}\green{1,}\blue{1,}\red{1,}\green{\text{-}1,}\blue{1}$ \\ $\overline{e}$ = -0.956(5) \\ $\overline{l}$ = 0.974(2) \\ $\overline{\mathcal{X}}$ = 0.00(2) 
        \vspace{0.2em}
        \end{tabular} & 
        \begin{tabular}{@{}c@{}} 
        $[0.92(2), 0.98(1), 0.99(1)], [0.96(2), 0.97(1), 0.98(1)], [0.93(2), 1.0, 0.94(2)], [0.96(2), 0.96(2), 0.87(3)], [0.92(2), 0.96(2), 1.0],$ \\ 
        $[0.92(2), 0.98(1), 0.96(2)], [0.94(2), 0.97(1), 0.98(1)], [0.90(3), 0.95(2), 0.95(2)], [0.96(2), 1.0, 0.99(1)]$ \\
        \arrayrulecolor{gray}\hline\arrayrulecolor{black}\vspace{0.2em}
        $0.974(7), 0.982(5), 0.964(7), 0.985(5), -0.980(6), 0.962(8)$ \\ 
        $-0.007(5), 0.203(5), -0.161(5), 0.051(2), -0.123(5), 0.045(3)$  
        \vspace{0.2em}
        \end{tabular} 
        \\ \hline
        
\begin{tabular} {@{}c@{}} $\red{1,}\green{1,}\blue{1,}\red{1,}\green{1,}\blue{\text{-}1}$ \\ $\overline{e}$ = -0.941(6) \\ $\overline{l}$ = 0.976(2) \\ $\overline{\mathcal{X}}$ = -0.01(2) 
        \vspace{0.2em}
        \end{tabular} & 
        \begin{tabular}{@{}c@{}} 
        $[0.92(2), 0.96(2), 0.97(1)], [0.89(3), 0.97(1), 0.96(2)], [0.95(2), 0.97(1), 0.98(1)], [0.88(3), 0.97(1), 0.98(1)], [0.92(2), 0.96(2), 0.96(2)],$ \\ 
        $[0.89(3), 0.92(2), 0.98(1)], [0.95(2), 0.96(2), 0.97(1)], [0.85(3), 0.93(2), 0.94(2)], [0.89(3), 0.96(2), 0.97(1)]$ \\
        \arrayrulecolor{gray}\hline\arrayrulecolor{black}\vspace{0.2em}
        $0.977(6), 0.974(6), 0.974(6), 0.975(7), 0.986(4), -0.972(6)$ \\ 
        $0.017(4), -0.037(4), 0.019(4), 0.049(2), -0.077(2), -0.040(4)$  
        \vspace{0.2em}
        \end{tabular} 
        \\ \hline
        
\begin{tabular} {@{}c@{}} $\red{\text{-}1,}\green{\text{-}1,}\blue{1,}\red{1,}\green{1,}\blue{1}$ \\ $\overline{e}$ = -0.934(6) \\ $\overline{l}$ = 0.968(3) \\ $\overline{\mathcal{X}}$ = 0.03(2) 
        \vspace{0.2em}
        \end{tabular} & 
        \begin{tabular}{@{}c@{}} 
        $[0.89(3), 0.96(2), 0.93(2)], [0.87(3), 0.96(2), 0.96(2)], [0.83(4), 0.92(2), 0.92(2)], [0.91(3), 0.96(2), 0.98(1)], [0.86(3), 0.99(1), 0.93(2)],$ \\ 
        $[0.97(1), 0.97(1), 0.94(2)], [0.86(3), 0.98(1), 0.99(1)], [0.86(3), 0.94(2), 0.98(1)], [0.93(2), 0.99(1), 0.96(2)]$ \\
        \arrayrulecolor{gray}\hline\arrayrulecolor{black}\vspace{0.2em}
        $-0.961(9), -0.970(7), 0.969(7), 0.974(7), 0.962(7), 0.974(6)$ \\ 
        $0.039(2), 0.041(2), 0.024(2), -0.088(4), 0.103(4), 0.046(5)$  
        \vspace{0.2em}
        \end{tabular} 
        \\ \hline
        
\begin{tabular} {@{}c@{}} $\red{1,}\green{\text{-}1,}\blue{\text{-}1,}\red{1,}\green{1,}\blue{1}$ \\ $\overline{e}$ = -0.930(6) \\ $\overline{l}$ = 0.967(3) \\ $\overline{\mathcal{X}}$ = 0.03(2) 
        \vspace{0.2em}
        \end{tabular} & 
        \begin{tabular}{@{}c@{}} 
        $[0.91(3), 0.98(1), 0.90(3)], [0.78(4), 0.95(2), 0.95(2)], [0.92(2), 0.98(1), 0.96(2)], [0.92(2), 0.96(2), 0.93(2)], [0.88(3), 0.90(3), 0.98(1)],$ \\ 
        $[0.89(3), 0.96(2), 0.94(2)], [0.93(2), 0.99(1), 0.96(2)], [0.85(3), 0.98(1), 0.94(2)], [0.81(4), 0.99(1), 0.99(1)]$ \\
        \arrayrulecolor{gray}\hline\arrayrulecolor{black}\vspace{0.2em}
        $0.973(6), -0.957(8), -0.971(7), 0.975(6), 0.952(8), 0.973(7)$ \\ 
        $-0.003(1), 0.052(2), 0.010(2), 0.021(4), -0.073(4), 0.162(4)$  
        \vspace{0.2em}
        \end{tabular} 
        \\ \hline
        
\begin{tabular} {@{}c@{}} $\red{\text{-}1,}\green{1,}\blue{\text{-}1,}\red{1,}\green{1,}\blue{1}$ \\ $\overline{e}$ = -0.928(7) \\ $\overline{l}$ = 0.971(2) \\ $\overline{\mathcal{X}}$ = 0.02(2) 
        \vspace{0.2em}
        \end{tabular} & 
        \begin{tabular}{@{}c@{}} 
        $[0.83(3), 0.95(2), 0.91(2)], [0.93(2), 0.95(2), 0.99(1)], [0.88(3), 0.96(2), 0.98(1)], [0.92(2), 0.98(1), 0.97(1)], [0.89(3), 0.95(2), 0.99(1)],$ \\ 
        $[0.86(3), 0.89(3), 0.94(2)], [0.86(3), 0.96(2), 0.98(1)], [0.84(3), 0.96(2), 0.95(2)], [0.87(3), 0.97(1), 0.94(2)]$ \\
        \arrayrulecolor{gray}\hline\arrayrulecolor{black}\vspace{0.2em}
        $-0.981(5), 0.971(6), -0.965(7), 0.990(4), 0.961(8), 0.959(8)$ \\ 
        $0.047(2), -0.003(1), -0.005(2), 0.020(4), -0.048(4), 0.121(4)$  
        \vspace{0.2em}
        \end{tabular} 
        \\ \hline
        
\begin{tabular} {@{}c@{}} $\red{1,}\green{1,}\blue{1,}\red{\text{-}1,}\green{\text{-}1,}\blue{1}$ \\ $\overline{e}$ = -0.927(7) \\ $\overline{l}$ = 0.965(3) \\ $\overline{\mathcal{X}}$ = -0.00(2) 
        \vspace{0.2em}
        \end{tabular} & 
        \begin{tabular}{@{}c@{}} 
        $[0.83(3), 0.91(3), 0.92(2)], [0.84(4), 0.98(1), 0.95(2)], [0.91(3), 0.98(1), 0.95(2)], [0.86(3), 0.95(2), 0.96(2)], [0.91(3), 0.93(2), 0.99(1)],$ \\ 
        $[0.85(3), 0.92(2), 0.93(2)], [0.94(2), 0.93(2), 0.99(1)], [0.88(3), 0.94(2), 0.99(1)], [0.92(2), 0.97(1), 0.96(2)]$ \\
        \arrayrulecolor{gray}\hline\arrayrulecolor{black}\vspace{0.2em}
        $0.977(6), 0.962(8), 0.959(8), -0.971(7), -0.955(9), 0.965(8)$ \\ 
        $-0.094(4), 0.095(5), 0.023(5), -0.042(2), 0.016(2), -0.023(1)$  
        \vspace{0.2em}
        \end{tabular} 
        \\ \hline
        
\end{tabular}}
    \egroup
    \end{center}\caption{Expectation values of the local, $Z$- and $X$-logical stabilizers for the first eleven of the 22 ground states of $D_4$ Hamiltonian (\ref{eq_hamiltonian}). $[e^\text{\ding{65}}_i,\ e^{\triangleleft}_i,\ e^{\triangleright}_i]$ denotes the expectation values of 12-body $A_s$ stabilizer ($e^\text{\ding{65}}_i$), left- ($e^{\triangleleft}_i$), and right-pointing ($e^{\triangleright}_i $) $B_t$ stabilizers. Plaquettes have been enumerated from left to right and top to bottom, and the order is given by index $i$. In the left column, we show the $Z$-logical signature $(\red{\mathcal{Z}_{RH}},\green{\mathcal{Z}_{GH}},\blue{\mathcal{Z}_{BH}},\red{\mathcal{Z}_{RV}},\green{\mathcal{Z}_{GV}},\blue{\mathcal{Z}_{BV}})$ of the ground state, energy density ($\overline{e}$), value of logical pinning function ($\overline{l}$), and the mean value of $X$-logical stabilizers ($\overline{\mathcal{X}}$).}\label{table:22_gss_a}\end {table*}

%% file: table-2.tex
    \begin {table*}
    \begin{center}
    	\bgroup
    	\def\arraystretch{1.2}% 
    	\scalebox{0.85}{
    		\begin{tabular}{c @{\ }@{\ }|@{\ }@{\ } c } 
    			%		\hline
    			Ground State &
    			
    			\begin{tabular}{@{}c@{}}
                $\qquad\qquad\qquad\qquad\qquad\qquad\qquad\qquad\qquad\quad[e^\text{\ding{65}}_1,\ e^{\triangleleft}_1,\ e^{\triangleright}_1],\ 
    				[e^\text{\ding{65}}_2,\ e^{\triangleleft}_2,\ e^{\triangleright}_2],\ \dots 
    				[e^\text{\ding{65}}_9,\ e^{\triangleleft}_9,\ e^{\triangleright}_9]
    				\qquad\qquad\qquad\qquad\qquad\qquad\qquad\qquad\qquad\quad
    				$ \vspace{0.2em}\\
    				\arrayrulecolor{gray}\hline\arrayrulecolor{black}\vspace{0.2em}
    				$
    				{\mathcal{Z}_{RH}},\ {\mathcal{Z}_{GH}},\ 
    				{\mathcal{Z}_{BH}},\ 
    				{\mathcal{Z}_{RV}},\ {\mathcal{Z}_{GV}},\ {\mathcal{Z}_{BV}}
    				$ \\ 
    				$
    				{\mathcal{X}_{RH}},\ {\mathcal{X}_{GH}},\ 
    				{\mathcal{X}_{BH}},\ 
    				{\mathcal{X}_{RV}},\ {\mathcal{X}_{GV}},\ {\mathcal{X}_{BV}}
    				$ 
    				\vspace{0.2em}
    			\end{tabular}
    			
    			\\ \hline\hline					
        
\begin{tabular} {@{}c@{}} $\red{1,}\green{1,}\blue{1,}\red{1,}\green{\text{-}1,}\blue{\text{-}1}$ \\ $\overline{e}$ = -0.919(7) \\ $\overline{l}$ = 0.959(3) \\ $\overline{\mathcal{X}}$ = 0.02(2) 
        \vspace{0.2em}
        \end{tabular} & 
        \begin{tabular}{@{}c@{}} 
        $[0.87(3), 0.96(2), 0.93(2)], [0.80(4), 0.97(1), 0.87(3)], [0.85(3), 0.99(1), 0.94(2)], [0.86(3), 0.96(2), 0.95(2)], [0.80(4), 0.81(4), 0.98(1)],$ \\ 
        $[0.82(4), 0.92(2), 0.96(2)], [0.95(2), 0.94(2), 0.99(1)], [0.91(3), 0.95(2), 0.99(1)], [0.90(3), 0.96(2), 1.0]$ \\
        \arrayrulecolor{gray}\hline\arrayrulecolor{black}\vspace{0.2em}
        $0.969(7), 0.94(1), 0.948(9), 0.978(6), -0.958(8), -0.958(8)$ \\ 
        $0.094(4), 0.007(4), 0.050(4), 0.054(1), -0.121(2), 0.029(2)$  
        \vspace{0.2em}
        \end{tabular} 
        \\ \hline
        
\begin{tabular} {@{}c@{}} $\red{1,}\green{1,}\blue{1,}\red{\text{-}1,}\green{1,}\blue{\text{-}1}$ \\ $\overline{e}$ = -0.940(7) \\ $\overline{l}$ = 0.971(3) \\ $\overline{\mathcal{X}}$ = -0.01(2) 
        \vspace{0.2em}
        \end{tabular} & 
        \begin{tabular}{@{}c@{}} 
        $[0.95(2), 0.99(1), 0.95(2)], [0.91(2), 0.990(9), 0.97(1)], [0.85(3), 0.95(2), 0.97(1)], [0.92(2), 0.95(2), 0.97(1)], [0.87(3), 0.91(2), 0.98(1)],$ \\ 
        $[0.94(2), 0.92(2), 0.97(1)], [0.90(3), 0.93(2), 0.99(1)], [0.89(3), 0.92(2), 0.93(2)], [0.93(2), 0.97(1), 0.97(1)]$ \\
        \arrayrulecolor{gray}\hline\arrayrulecolor{black}\vspace{0.2em}
        $0.975(7), 0.961(7), 0.966(8), -0.985(4), 0.973(7), -0.966(8)$ \\ 
        $0.074(4), -0.102(4), 0.052(4), 0.021(2), -0.056(1), -0.065(2)$  
        \vspace{0.2em}
        \end{tabular} 
        \\ \hline
        
\begin{tabular} {@{}c@{}} $\red{\text{-}1,}\green{1,}\blue{1,}\red{\text{-}1,}\green{1,}\blue{1}$ \\ $\overline{e}$ = -0.947(6) \\ $\overline{l}$ = 0.974(2) \\ $\overline{\mathcal{X}}$ = -0.00(2) 
        \vspace{0.2em}
        \end{tabular} & 
        \begin{tabular}{@{}c@{}} 
        $[0.85(3), 0.95(2), 0.92(2)], [0.90(3), 0.95(2), 0.96(2)], [0.91(2), 0.96(2), 0.99(1)], [0.94(2), 0.96(2), 0.96(2)], [0.93(2), 0.95(2), 0.99(1)],$ \\ 
        $[0.96(1), 0.96(1), 0.97(1)], [0.91(2), 1.0, 0.98(1)], [0.90(3), 0.97(1), 0.96(1)], [0.90(3), 0.97(1), 0.97(1)]$ \\
        \arrayrulecolor{gray}\hline\arrayrulecolor{black}\vspace{0.2em}
        $-0.976(6), 0.979(6), 0.970(8), -0.976(7), 0.972(6), 0.970(7)$ \\ 
        $-0.047(4), 0.069(2), 0.083(2), -0.057(4), -0.076(2), 0.012(2)$  
        \vspace{0.2em}
        \end{tabular} 
        \\ \hline
        
\begin{tabular} {@{}c@{}} $\red{1,}\green{\text{-}1,}\blue{1,}\red{1,}\green{\text{-}1,}\blue{1}$ \\ $\overline{e}$ = -0.941(6) \\ $\overline{l}$ = 0.967(3) \\ $\overline{\mathcal{X}}$ = 0.05(2) 
        \vspace{0.2em}
        \end{tabular} & 
        \begin{tabular}{@{}c@{}} 
        $[0.87(3), 0.94(2), 0.91(2)], [0.89(3), 0.92(2), 0.90(3)], [0.89(3), 0.99(1), 0.96(2)], [0.92(2), 1.0, 0.93(2)], [0.94(2), 1.0, 0.99(1)],$ \\ 
        $[0.90(3), 0.93(2), 0.95(2)], [0.95(2), 0.99(1), 0.97(1)], [0.90(3), 0.94(2), 0.99(1)], [0.90(3), 0.99(1), 0.97(1)]$ \\
        \arrayrulecolor{gray}\hline\arrayrulecolor{black}\vspace{0.2em}
        $0.984(5), -0.968(8), 0.954(8), 0.975(6), -0.968(8), 0.951(9)$ \\ 
        $0.027(2), 0.053(5), 0.029(3), 0.034(2), 0.190(5), -0.029(2)$  
        \vspace{0.2em}
        \end{tabular} 
        \\ \hline
        
\begin{tabular} {@{}c@{}} $\red{1,}\green{1,}\blue{\text{-}1,}\red{1,}\green{1,}\blue{\text{-}1}$ \\ $\overline{e}$ = -0.923(7) \\ $\overline{l}$ = 0.963(3) \\ $\overline{\mathcal{X}}$ = 0.04(2) 
        \vspace{0.2em}
        \end{tabular} & 
        \begin{tabular}{@{}c@{}} 
        $[0.96(2), 0.99(1), 0.97(1)], [0.80(4), 0.91(3), 0.96(2)], [0.91(3), 0.98(1), 0.94(2)], [0.83(4), 0.94(2), 0.99(1)], [0.89(3), 0.88(3), 0.97(1)],$ \\ 
        $[0.85(3), 0.96(2), 0.94(2)], [0.96(2), 0.97(1), 0.99(1)], [0.85(3), 0.93(2), 0.88(3)], [0.80(4), 0.93(2), 0.94(2)]$ \\
        \arrayrulecolor{gray}\hline\arrayrulecolor{black}\vspace{0.2em}
        $0.967(7), 0.951(9), -0.981(5), 0.95(1), 0.957(8), -0.970(7)$ \\ 
        $-0.002(2), 0.106(2), 0.053(5), 0.058(2), 0.095(3), -0.053(5)$  
        \vspace{0.2em}
        \end{tabular} 
        \\ \hline
        
\begin{tabular} {@{}c@{}} $\red{\text{-}1,}\green{\text{-}1,}\blue{\text{-}1,}\red{1,}\green{1,}\blue{1}$ \\ $\overline{e}$ = -0.935(7) \\ $\overline{l}$ = 0.972(2) \\ $\overline{\mathcal{X}}$ = 0.01(2) 
        \vspace{0.2em}
        \end{tabular} & 
        \begin{tabular}{@{}c@{}} 
        $[0.85(3), 0.95(2), 0.91(2)], [0.84(4), 0.96(2), 0.92(2)], [0.93(2), 0.97(1), 0.97(1)], [0.83(4), 0.94(2), 0.96(2)], [0.95(2), 0.95(2), 0.97(1)],$ \\ 
        $[0.90(3), 0.95(2), 0.99(1)], [0.90(3), 0.98(1), 0.97(1)], [0.87(3), 0.96(2), 0.99(1)], [0.89(3), 0.99(1), 0.98(1)]$ \\
        \arrayrulecolor{gray}\hline\arrayrulecolor{black}\vspace{0.2em}
        $-0.971(7), -0.975(6), -0.971(7), 0.978(6), 0.972(6), 0.963(7)$ \\ 
        $0.012(1), 0.060(1), 0.012(1), -0.077(4), 0.041(5), -0.005(4)$  
        \vspace{0.2em}
        \end{tabular} 
        \\ \hline
        
\begin{tabular} {@{}c@{}} $\red{1,}\green{1,}\blue{1,}\red{\text{-}1,}\green{\text{-}1,}\blue{\text{-}1}$ \\ $\overline{e}$ = -0.938(6) \\ $\overline{l}$ = 0.970(2) \\ $\overline{\mathcal{X}}$ = 0.00(2) 
        \vspace{0.2em}
        \end{tabular} & 
        \begin{tabular}{@{}c@{}} 
        $[0.92(2), 0.96(2), 0.97(1)], [0.93(2), 0.94(2), 0.98(1)], [0.84(3), 0.99(1), 0.96(2)], [0.96(2), 0.98(1), 0.98(1)], [0.89(3), 0.95(2), 0.96(2)],$ \\ 
        $[0.86(3), 0.96(2), 0.95(2)], [0.80(4), 0.91(3), 0.97(1)], [0.88(3), 0.97(1), 0.90(3)], [0.97(1), 0.99(1), 0.98(1)]$ \\
        \arrayrulecolor{gray}\hline\arrayrulecolor{black}\vspace{0.2em}
        $0.981(5), 0.952(8), 0.975(6), -0.974(7), -0.966(7), -0.973(6)$ \\ 
        $0.093(4), -0.003(5), -0.081(4), 0.061(1), -0.062(1), 0.021(1)$  
        \vspace{0.2em}
        \end{tabular} 
        \\ \hline
        
\begin{tabular} {@{}c@{}} $\red{\text{-}1,}\green{\text{-}1,}\blue{1,}\red{\text{-}1,}\green{\text{-}1,}\blue{1}$ \\ $\overline{e}$ = -0.937(7) \\ $\overline{l}$ = 0.974(2) \\ $\overline{\mathcal{X}}$ = -0.01(2) 
        \vspace{0.2em}
        \end{tabular} & 
        \begin{tabular}{@{}c@{}} 
        $[0.86(3), 0.92(2), 0.92(2)], [0.87(3), 0.96(2), 0.94(2)], [0.87(3), 0.95(2), 0.97(1)], [0.92(2), 1.0, 0.93(2)], [0.90(3), 0.96(2), 0.98(1)],$ \\ 
        $[0.89(3), 0.94(2), 0.98(1)], [0.93(2), 0.98(1), 1.0], [0.92(2), 0.96(2), 0.99(1)], [0.88(3), 0.98(1), 0.93(2)]$ \\
        \arrayrulecolor{gray}\hline\arrayrulecolor{black}\vspace{0.2em}
        $-0.986(4), -0.981(6), 0.965(7), -0.988(4), -0.970(7), 0.956(8)$ \\ 
        $0.005(2), 0.012(2), -0.059(1), 0.058(2), -0.023(2), -0.073(1)$  
        \vspace{0.2em}
        \end{tabular} 
        \\ \hline
        
\begin{tabular} {@{}c@{}} $\red{1,}\green{\text{-}1,}\blue{\text{-}1,}\red{1,}\green{\text{-}1,}\blue{\text{-}1}$ \\ $\overline{e}$ = -0.921(7) \\ $\overline{l}$ = 0.966(3) \\ $\overline{\mathcal{X}}$ = 0.00(2) 
        \vspace{0.2em}
        \end{tabular} & 
        \begin{tabular}{@{}c@{}} 
        $[0.83(4), 0.91(3), 0.98(1)], [0.93(2), 1.0, 0.94(2)], [0.88(3), 0.92(2), 0.99(1)], [0.88(3), 0.97(1), 0.94(2)], [0.75(4), 0.83(3), 0.95(2)],$ \\ 
        $[0.92(2), 0.94(2), 0.95(2)], [0.84(3), 0.94(2), 0.98(1)], [0.86(3), 0.96(2), 0.95(2)], [0.93(2), 0.99(1), 0.95(2)]$ \\
        \arrayrulecolor{gray}\hline\arrayrulecolor{black}\vspace{0.2em}
        $0.986(4), -0.948(9), -0.965(7), 0.984(5), -0.957(8), -0.953(8)$ \\ 
        $0.005(1), 0.035(2), -0.054(2), 0.044(1), 0.014(2), -0.036(2)$  
        \vspace{0.2em}
        \end{tabular} 
        \\ \hline
        
\begin{tabular} {@{}c@{}} $\red{\text{-}1,}\green{1,}\blue{\text{-}1,}\red{\text{-}1,}\green{1,}\blue{\text{-}1}$ \\ $\overline{e}$ = -0.907(8) \\ $\overline{l}$ = 0.964(3) \\ $\overline{\mathcal{X}}$ = 0.02(1) 
        \vspace{0.2em}
        \end{tabular} & 
        \begin{tabular}{@{}c@{}} 
        $[0.90(3), 0.94(2), 0.88(3)], [0.75(4), 0.95(2), 0.92(2)], [0.80(4), 0.89(3), 0.97(1)], [0.84(3), 0.97(1), 0.92(2)], [0.91(3), 0.89(3), 0.98(1)],$ \\ 
        $[0.87(3), 0.86(3), 0.92(2)], [0.84(3), 0.96(2), 1.0], [0.89(3), 0.96(1), 0.93(2)], [0.76(4), 1.0, 0.99(1)]$ \\
        \arrayrulecolor{gray}\hline\arrayrulecolor{black}\vspace{0.2em}
        $-0.988(4), 0.963(7), -0.959(8), -0.988(4), 0.93(1), -0.954(8)$ \\ 
        $0.053(2), 0.026(1), 0.135(2), 0.023(2), -0.043(1), -0.076(2)$  
        \vspace{0.2em}
        \end{tabular} 
        \\ \hline
        
\begin{tabular} {@{}c@{}} $\red{\text{-}1,}\green{\text{-}1,}\blue{\text{-}1,}\red{\text{-}1,}\green{\text{-}1,}\blue{\text{-}1}$ \\ $\overline{e}$ = -0.863(9) \\ $\overline{l}$ = 0.930(4) \\ $\overline{\mathcal{X}}$ = 0.00(1) 
        \vspace{0.2em}
        \end{tabular} & 
        \begin{tabular}{@{}c@{}} 
        $[0.78(4), 0.85(3), 0.80(4)], [0.89(3), 0.99(1), 0.89(3)], [0.84(3), 0.90(3), 1.0], [0.89(3), 0.96(2), 0.84(4)], [0.78(4), 0.74(4), 0.98(1)],$ \\ 
        $[0.60(5), 0.76(4), 0.86(3)], [0.78(4), 0.95(2), 0.98(1)], [0.77(4), 0.94(2), 0.87(3)], [0.74(4), 0.96(2), 0.97(1)]$ \\
        \arrayrulecolor{gray}\hline\arrayrulecolor{black}\vspace{0.2em}
        $-0.986(4), -0.91(1), -0.92(1), -0.980(5), -0.89(1), -0.89(1)$ \\ 
        $0.038(1), 0.081(1), -0.108(1), 0.045(1), 0.005(1), -0.046(1)$  
        \vspace{0.2em}
        \end{tabular} 
        \\ \hline
        
\end{tabular}}
    \egroup
    \end{center}\caption{Expectation values of the local, $Z$- and $X$-logical stabilizers for the remaining eleven of the 22 ground states of $D_4$ Hamiltonian (\ref{eq_hamiltonian}). }\label{table:22_gss_b}\end {table*}